\documentclass[prd,preprint,tightenlines,floatfix,
nofootinbib,eqsecnum,superscriptaddress]{revtex4-1}

\usepackage[T1]{fontenc}		

\usepackage{amsmath,amsfonts,amssymb,amstext,mathrsfs}
\usepackage{mathpazo}

\usepackage[dvips]{graphicx}
\usepackage{epsf,float}
\usepackage{revsymb}

\usepackage{dcolumn}
\usepackage{braket}
\usepackage{color,xcolor}
\usepackage{graphicx}
\usepackage{subfigure}
\usepackage{multirow}
\usepackage{tabularx}
\usepackage{pstricks}
\usepackage[section]{placeins}
\usepackage{booktabs}
\usepackage{array}

\usepackage{hyperref}

\newcommand{\Pom}{\mathbb{P}}

\newcommand{\bq}{\mbox{\boldmath $q$}}
\newcommand{\bqa}{\mbox{\boldmath $q$}_{1}}
\newcommand{\bqb}{\mbox{\boldmath $q$}_{2}}

\newcommand{\bp}{\mbox{\boldmath $p$}}

\begin{document}


\title{\boldmath 
Production of $f_{0}(980)$ meson at the LHC:\\
Color evaporation versus color-singlet gluon-gluon fusion}

\vspace{0.6cm}

\author{Piotr Lebiedowicz}
 \email{Piotr.Lebiedowicz@ifj.edu.pl}
\affiliation{Institute of Nuclear Physics Polish Academy of Sciences, ul. Radzikowskiego 152, PL-31342 Krak{\'o}w, Poland}

\author{Rafa{\l} Maciu{\l}a}
\email{Rafal.Maciula@ifj.edu.pl}
\affiliation{Institute of Nuclear Physics Polish Academy of Sciences, ul. Radzikowskiego 152, PL-31342 Krak{\'o}w, Poland}

\author{Antoni Szczurek
\footnote{Also at \textit{College of Natural Sciences, 
Institute of Physics, University of Rzesz{\'o}w, 
ul. Pigonia 1, PL-35310 Rzesz{\'o}w, Poland}.}}
\email{Antoni.Szczurek@ifj.edu.pl}
\affiliation{Institute of Nuclear Physics Polish Academy of Sciences, ul. Radzikowskiego 152, PL-31342 Krak{\'o}w, Poland}

\begin{abstract}
The production of the $f_{0}(980)$ meson at high energies is not well 
understood.
We investigate two different potential mechanisms for inclusive scalar 
meson production in the $k_t$-factorization approach: color-singlet 
gluon-gluon fusion and color evaporation model. 
The $\gamma^* \gamma^* \to f_0(980)$ form factor(s)
can be constraint from the $f_0(980)$ radiative decay width.
The $g^* g^* \to f_0(980)$ form factors are obtained by
a replacement of $\alpha_{\rm{em}}$ electromagnetic coupling constant 
by $\alpha_{\rm{s}}$ strong coupling constant 
and appropriate
color factors. The form factors for the two couplings are parametrized
with a function motivated by recent results for scalar quarkonia.
The differential cross sections are calculated in the
$k_t$-factorization approach with modern unintegrated gluon distributions.
Unlike for quarkonia it seems rather difficult to describe a 
preliminary ALICE data for inclusive production of $f_0(980)$ exclusively
by the color singlet gluon-gluon fusion mechanism.
Two different scenarios for flavor structure of $f_0(980)$ are
considered in this context.
We consider also mechanism of fusion of quark-antiquark
associated with soft gluon emission in a
phenomenological color evaporation model (CEM) used sometimes for 
quarkonium production. 
Here we use $k_t$-factorization version of CEM to include 
higher-order contributions.
In addition, for comparison we consider also NLO  collinear 
approach with $q \bar{q} q$ and $q \bar{q} g$ 
color octet partonic final states.
Both approaches lead to a similar result. 
However, very large probabilities are required to describe 
the preliminary ALICE data.
The pomeron-pomeron fusion mechanism is also discussed and results
are quantified.
\end{abstract} 

\maketitle

\section{Introduction}

The production of light mesons in high-energy proton-proton collisions
is rather poorly understood. 
Representative examples are production of 
$f_{0}(500)$, $\rho(770)$, $f_{0}(980)$ or $f_{2}(1270)$. 
Parallel we discussed inclusive production of 
$f_2(1270)$ meson in proton-proton collisions 
\cite{LMS_2020} where it is found that the preliminary 
ALICE data \cite{Lee:thesis} can be almost explained 
at higher $f_2(1270)$ transverse momentum ($p_t > 3$~GeV) 
using color-singlet gluon-gluon fusion mechanism.
The $f_2(1270)$ meson is usually considered to have a 
$\frac{1}{\sqrt{2}} \left( u \bar u + d \bar d \right)$ flavor structure.
Here we wish to explore the situation for the production of a rather
enigmatic $f_0(980)$.
 
In general, light scalar mesons are poorly understood \cite{Close:2002zu}.
In particular, it is not clear whether they are of the $q \bar q$
character or are tetraquarks \cite{Maiani:2004uc}.
Most mesons are thought to be formed from combinations of $q \bar{q}$.
In the literature, the hadronic structure of the $f_{0}(980)$ meson
has been discussed for decades and there are many different interpretations,
from the conventional $q \bar{q}$ picture 
\cite{Tornqvist:1995kr,Boglione:2002vv}
to multiquark \cite{Jaffe:1976ig,Jaffe:1976ih} or $K \bar{K}$ bound states 
\cite{Weinstein:1982gc,Weinstein:1983gd,Weinstein:1990gu,Baru:2003qq}.
Some authors introduce the concept of $q q \bar{q} \bar{q}$ states
\cite{Maiani:2004uc} or even superpositions of the tetraquark state 
with the $q \bar{q}$ state \cite{Hooft:2008we,Fleischer:2011au}.
The structure of $f_0(980)$ can be studied also in nonsemileptonic 
decays of $D, D_s$ mesons \cite{Maiani:2007iw} or $B, B_s$ mesons 
\cite{Cheng:2005nb,Stone:2013eaa}.

Note that $f_{0}(980)$ state was seen in both $\pi \pi$ and $K \bar{K}$
channel \cite{Tanabashi:2018oca} with a considerable branching fraction.
For the branching ratios see the discussion, e.g., in 
Refs.~\cite{Fleischer:2011au,Aaij:2014siy}.

In the present letter we investigate whether the gluon-gluon fusion
or color evaporation approaches known from quarkonium production
can explain the new preliminary ALICE data \cite{Lee:thesis}.
As this is a first analysis on the subject we shall consider a 
simple $q \bar q$ structure of $f_0(980)$ meson. 
We shall consider different flavor combinations.
This has of course important consequences for 
$\gamma^* \gamma^* \to f_0(980)$ coupling due to charges of quarks/antiquarks.
Such couplings are important ingredients for calculating $f_0(980)$ contribution to light-by-light component to anomalous magnetic moment of the muon \cite{Pauk:2014rta,Colangelo:2014dfa,Dorokhov:2015psa}.
In Ref.~\cite{DeFazio:2001uc} it was argued that $f_0(980)$ must be
dominantly $s \bar s$ to describe radiative decay $\phi \to f_0(980) \gamma$. 
This is dictated by the fact that
$\Gamma(\phi \to f_0(980) \gamma) \gg \Gamma(\phi \to a_0(980) \gamma)$. 
In Ref.~\cite{Kroll:2016mbt} the $\gamma^* - f_0(980)$ transition
form factor was studied assuming the simple $s \bar{s}$ structure.
Only $F_{TT}$ transverse form factor was included in this analysis.
The role of $F_{LL}$ longitudinal form factor was not studied so far.

\section{Some details of the model calculations}
\label{sec:model}

\subsection{The $\gamma^* \gamma^* \to f_0(980)$ fusion process}

In the formalism presented e.g. in \cite{Pascalutsa:2012pr} 
the covariant matrix element 
for the $\gamma^* \gamma^* \to f_0(980)$ process 
is written as:
\begin{eqnarray}
{\cal M}^{\mu \nu} &=& 4 \pi \alpha_{\rm em}
\left. 
\frac{\nu}{m_{f_{0}}}
\Biggl[ -R^{\mu \nu}(q_1,q_2)\,
F_{TT}(Q_1^2,Q_2^2) 
\right. \nonumber \\
&& 
\left. 
+ \frac{\nu}{X}
\left(q_1^{\mu} + \frac{Q_1^2}{\nu} q_2^{\mu} \right)
\left(q_2^{\nu} + \frac{Q_2^2}{\nu} q_1^{\nu} \right)\,
F_{LL}(Q_1^2,Q_2^2) \Biggr] 
\right.\,,
\label{matrix_element}    
\end{eqnarray}
where 
$\nu = (q_{1} \cdot q_{2})$, $X = \nu^{2} - q_1^2 q_2^2 $, and
\begin{equation}
R^{\mu \nu}(q_{1},q_{2}) = -g^{\mu \nu} + \frac{1}{X}
\left[
\nu \left( q_1^{\mu} q_2^{\nu} + q_2^{\mu} q_1^{\nu} \right)
- q_1^2 q_2^{\mu} q_2^{\nu} - q_2^2 q_1^{\mu} q_1^{\nu}
\right]\,.
\end{equation}
Here $q_{1}$ and $q_{2}$ denote the momenta of the photons,
$Q_{1}^{2} = -q_{1}^{2}$, $Q_{2}^{2} = -q_{2}^{2}$, 
and $m_{f_{0}}$ is mass of the $f_{0}(980)$ meson.
In Eq.~(\ref{matrix_element}), the scalar meson structure information
is encoded in the form factors $F_{TT}$ and $F_{LL}$
which are functions of the virtualities of both photons.
$F_{TT}$ or $F_{LL}$ correspond to the situation where
either both photons are transverse or longitudinal, respectively.
By definition the form factors are dimensionless.

For scalar quarkonium states a microscopic calculation is reliable; 
see \cite{Babiarz:2020jkh}.
For light mesons the situation is more complicated.
Here we will try to rather parametrize the form factors.

The two-photon decay width of the $f_0(980)$ meson
can be calculated as:
\begin{equation}
\Gamma(f_{0}(980) \to \gamma \gamma) = \frac{\pi \alpha_{\rm{em}}^2}{4}
m_{f_{0}} |F_{TT}(0,0)|^2 \,.
\label{radiative_decay_width}
\end{equation}
Only $F_{TT}$ form factor can be constraint from (\ref{radiative_decay_width}).
The radiative decay width is relatively well known, 
see \cite{Tanabashi:2018oca}. 
Using the average decay width quoted in \cite{Tanabashi:2018oca}
\begin{equation}
\Gamma(f_{0}(980) \to \gamma \gamma) = 0.31 \; {\rm keV} \,,
\label{radiative_decay_width_f0980}
\end{equation}
and $m_{f_{0}} = 980$~MeV 
we obtain from (\ref{radiative_decay_width})
$|F_{TT}(0,0)| = 0.087$.
Then the transverse form factor is parametrized as:
\begin{eqnarray}
\frac{F_{TT}(Q_1^2,Q_2^2)}{F_{TT}(0,0)} &=& 
\left( \frac{\Lambda_M^2}{Q_1^2+Q_2^2+\Lambda_M^2} \right) \,,
\label{monopole} \\
\frac{F_{TT}(Q_1^2,Q_2^2)}{F_{TT}(0,0)} &=& 
\left( \frac{\Lambda_D^2}{Q_1^2+Q_2^2+\Lambda_D^2} \right)^2 \,, 
\label{dipole}           
\end{eqnarray}
where cut-off parameters 
$\Lambda_M$ or $\Lambda_D$ are expected
to be of order of 1~GeV.
Both monopole (\ref{monopole}) and dipole (\ref{dipole}) parametrizations 
of $F_{TT}$ will be used in the following.
In the calculations we take $\Lambda_M = \Lambda_D = m_{f_{0}}$.

The $F_{LL}$ form factor is rather unknown but via construction
do not enter the formula for the radiative decay width (\ref{radiative_decay_width})
as
\begin{equation}
F_{LL}(0,Q_2^2) = F_{LL}(Q_1^2,0) = 0 \,.
\label{FLL_condition}
\end{equation}
We propose to use the following parametrization for the $F_{LL}$
form factor:
\begin{eqnarray}
F_{LL}(Q_1^2,Q_2^2) = R_{LL/TT} \,
\frac{Q_1^2}{M_0^2 + Q_1^2}\,
\frac{Q_2^2}{M_0^2 + Q_2^2}\,
F_{TT}(Q_1^2,Q_2^2)  \,.
\label{F_LL_parametrization}
\end{eqnarray}
Such a form is consistent with a microscopic calculation
for $\gamma^* \gamma^* \to \chi_{c0}$ \cite{Babiarz:2020jkh}
using quarkonium wave functions obtained from the potential models.
In our present case we expect $R_{LL/TT} \approx \pm 0.5$ 
and $M_0 \sim m_{f_{0}}$.

\subsection{Color singlet $g^* g^* \to f_{0}(980)$ fusion}

\begin{figure}[!ht]
\begin{center}
\includegraphics[width=0.35\textwidth]{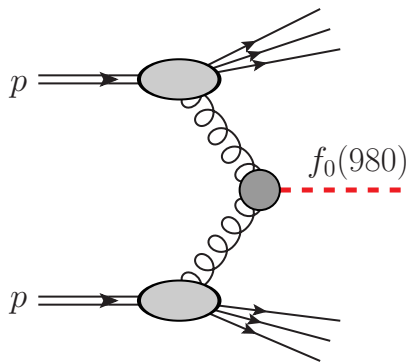}
\caption{\label{fig:diagram_gg_f0}
\small
General diagram for inclusive $f_0(980)$ production
via gluon-gluon fusion in proton-proton collisions.}
\end{center}
\end{figure}

In Fig.~\ref{fig:diagram_gg_f0} we show a generic Feynman diagram for 
$f_0(980)$ meson production in proton-proton collision via 
gluon-gluon fusion.
This diagram illustrates the situation adequate for 
the $k_t$-factorization calculations used in the present paper.

The differential cross section for inclusive $f_0(980)$ meson production 
via the $g^* g^* \to f_0(980)$ fusion in the $k_t$-factorization approach
can be written as:
\begin{eqnarray}
{d \sigma \over dy d^2\bp} = \int {d^2 \bqa \over \pi \bqa^2} 
{\cal F}_{g}(x_1,\bqa^2) \int {d^2 \bqb \over \pi \bqb^2}  
{\cal F}_{g}(x_2,\bqb^2) \, \delta^{(2)} (\bqa + \bqb - \bp ) \, 
{\pi \over (x_1 x_2 s)^2} \overline{|{\cal{M}}|^2} \,.
\label{eq:cross_section_2}
\end{eqnarray}
Here $\bqa$, $\bqb$ and $\bp$ denote the transverse momenta of the
gluons and the $f_{0}(980)$ meson.
${\cal M}_{g^{*} g^{*} \to f_0}$ is the off-shell matrix element 
for the hard subprocess and ${\cal F}_{g}$ are the gluon unintegrated 
distribution functions (UGDFs) for both colliding protons. 
The UGDFs depend on gluon longitudinal momentum fractions
$x_{1,2} = m_{T} \exp(\pm {\rm y})/\sqrt{s}$
and $\bqa^2, \bqb^2$ entering the hard process.
In principle, they can depend also on factorization scales 
$\mu_{F,i}^2$, $i = 1, 2$.
It is reasonable to assume $\mu_{F,1}^2 = \mu_{F,2}^2 = m_{T}^2$.
Here $m_{T}$ is transverse mass of the produced $f_0(980)$ meson;
$m_{T} = \sqrt{\bp^{2} + m_{f_{0}}^{2}}$.
The~$\delta^{(2)}$ function in Eq.~(\ref{eq:cross_section_2}) 
can be easily eliminated by introducing
$\bqa + \bqb$ and $\bqa - \bqb$ transverse momenta \cite{Cisek:2017gno}.

The off-shell matrix element can be written as (we restore the color
indices $a$ and $b$)
\begin{equation}
{\cal{M}}^{ab} = {q_{1t}^\mu q_{2t}^\nu \over |\bqa| |\bqb|}
{\cal{M}}^{ab}_{\mu \nu}  = 
{q_{1+} q_{2-} \over |\bqa| |\bqb|} n^{+\mu} n^{-\nu}
{\cal{M}}^{ab}_{\mu \nu} = 
{x_1 x_2 s \over 2 |\bqa| |\bqb| } n^{+\mu} n^{-\nu} {\cal{M}}^{ab}_{\mu \nu}
\end{equation}
with the lightcone components of gluon momenta
$q_{1+} = x_{1} \sqrt{s/2}$, $q_{2-} = x_{2} \sqrt{s/2}$.

The $g^* g^* \to f_0(980)$ coupling entering in the matrix element squared
can be obtained from that for $\gamma^* \gamma^* \to f_0(980)$ coupling 
(see e.g. \cite{Babiarz:2019mag}) 
by the following replacement:
\begin{equation}
\alpha_{\rm{em}}^2 \to \alpha_{\rm s}^2  \,
\frac{1}{4 N_c (N_c^2 - 1)} \,
\frac{1}{(<e_q^2>)^2} \,.
\label{replacement}
\end{equation}
$(<e_q^2>)$ above strongly depends on the flavor
structure of the wave function.  In the following we consider 
a few examples of quark-flavor composition:
\begin{eqnarray}
&&
\bullet \quad \ket{f_{0}(980)} = 
\frac{1}{\sqrt{2}} \left( \ket{u \bar{u}} + \ket{d \bar{d}} \right)\,, 
\label{nnbar}\\
&&
\bullet \quad \ket{f_{0}(980)} = 
\ket{s \bar{s}}\,,
\label{ssbar}\\
&&
\bullet \quad \ket{f_{0}(980)} = 
  \frac{1}{\sqrt{2}} \left( \ket{[su] [\bar{s}\bar{u}]} + \ket{[sd] [\bar{s}\bar{d}]} \right)\,.
\label{tetraquark}
\end{eqnarray}
The first function is written in analogy to the rather well known
flavor wave function of $f_2(1270)$ meson. The second function
was suggested by analysis of radiative decays of $\phi$ meson as
discussed in the introduction. The last function 
(tetraquark)
is supported by spectroscopy of scalar mesons 
(see e.g.~\cite{Maiani:2004uc}).
The scalar mesons with masses below 1~GeV can be understood to be
of the tetraquark character and those above 1~GeV as of the $q \bar{q}$
or glueball character. There is, however, no general consensus
and the situation is open in our opinion. To reach final picture 
one must include very different processes simultaneously.

In realistic calculations the running of strong coupling constants
must be included.
In our numerical calculations presented below, we set the factorization 
scale to $\mu_F^2 = m_{T}^2$, and the renormalization scale is taken 
in the form:
\begin{equation}
\alpha_{\rm s}^2 \to 
\alpha_{\rm s}(\max{\{m_{T}^2,\bqa^2\}})\,
\alpha_{\rm s}(\max{\{m_{T}^2,\bqb^2\}})\,.
\label{alpha_s}
\end{equation}
%

\subsection{Color evaporation model (CEM)}

\begin{figure}[!ht]
\begin{center}
\includegraphics[width=0.4\textwidth]{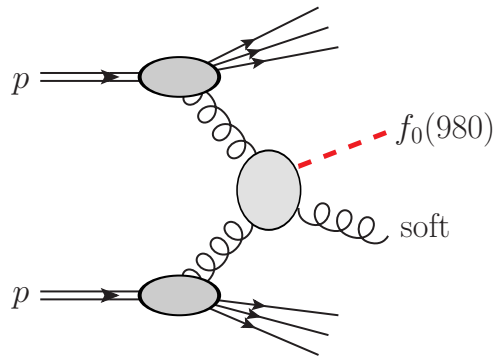}
\caption{\label{fig:diagram_CEM}
\small
General diagram for inclusive $f_0(980)$ production
in proton-proton collisions in the color evaporation approach.}
\end{center}
\end{figure}

The general diagram representing the color evaporation model (CEM) \cite{Fritzsch:1977ay,Halzen:1977rs} is shown
in Fig.~\ref{fig:diagram_CEM}. 
In this approach one uses 
the perturbative calculation of $q\bar q$ minijets.

\begin{figure}[!ht]
\begin{center}
\includegraphics[width=0.35\textwidth]{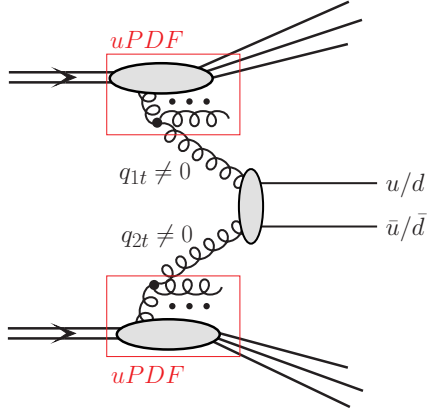}
\caption{\label{fig:diagram_gg_qqbar}
\small
Typical $k_{t}$-factorization process with the production of 
$u \bar{u}$ and $d \bar{d}$ pairs
that are intermediate state for color evaporation.}
\end{center}
\end{figure}

Fig.~\ref{fig:diagram_gg_qqbar} represents 
diagram with $q \bar{q}$ production in the $k_t$-factorization approach 
in proton-proton collisions.
Here, we calculate $u \bar u$ and $d \bar d$ production, 
or alternatively $s \bar s$ production,
in a similar way as it was done for
$c \bar{c}$ production \cite{Maciula:2018bex}.
The color of the $u \bar{u}$ or $d \bar{d}$ is typically 
in the octet representation.
The further emissions of soft gluons are not explicit but will be
contained in a multiplicative factor ${\rm P_{CEM}}$ defined below.

Everything is contained in a suitable renormalization of the $q\bar q$-cross section 
when integrating over certain limits in the $q \bar{q}$ invariant mass. 
Having calculated differential cross section for $q\bar q$-pair
production one can obtain the cross section for $f_{0}(980)$ meson
within the framework of the CEM. 
The $q\bar q \to f_{0}(980)$ transition can be
formally written as follows:
\begin{equation}
\frac{d\sigma_{f_{0}}(p_{f_{0}})}{d^3 p_{f_{0}}} = 
{\rm P_{CEM}} \int_{m_{f_{0}}-\Delta M}^{m_{f_{0}}+\Delta M} d^3 P_{q\bar q} \; d M_{q\bar q} \frac{d\sigma_{q\bar q}(M_{q\bar q},P_{q\bar q})}{dM_{q\bar q} \,d^3 P_{q\bar q}} 
\delta^3(\vec{p}_{f_{0}}-\frac{m_{f_{0}}}{M_{q\bar q}} \vec{P}_{q \bar q})\,,
\label{transition}
\end{equation}
where ${\rm P_{CEM}}$ is the probability of the $q\bar q \to f_{0}(980)$ transition which is fitted to the experimental data, 
$M_{q\bar q}$ and $P_{q\bar q} = |\vec{P}_{q\bar q}|$ are the invariant mass and momentum
of the $q\bar q$ system. 
Here we take $\Delta M = 100$ MeV.

\begin{figure}[!ht]
\begin{center}
\includegraphics[width=0.35\textwidth]{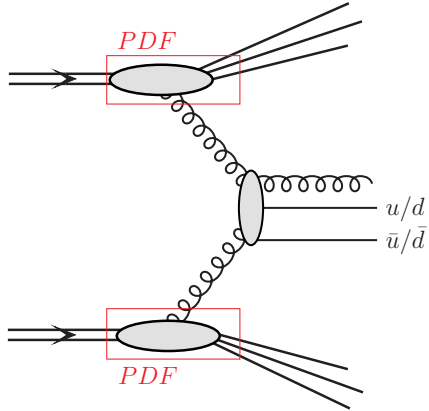}
\caption{\label{fig:diagram_qqbarg}
\small
An alternative collinear approach with the production of $u \bar u$ and $d \bar d$ 
pairs associated with soft gluon emission
that are intermediate state for color evaporation.}
\end{center}
\end{figure}

In Fig.\ref{fig:diagram_qqbarg} we show an example of the diagram
relevant for collinear next-to-leading order approach.
A full list of processes included in the calculation will be presented
in the result section.
Within the collinear-factorization approach in the leading-order (LO) approximation, the transverse momentum of the $q \bar q$ pair is equal to zero. In fact, the NLO diagrams for the inclusive minijets, 
such as $gg \to g q\bar q$ or $qg \to q q\bar q$, constitute the LO contributions for the $q\bar q$-pair transverse momentum. Similarly, the next-to-next-to-leading-order (NNLO) topologies for this quantity are effectively NLO. The situation is different in the $k_{t}$-factorization approach where nonzero $q \bar q$-pair transverse momentum can be obtained already at leading order within the $g^*g^* \to q \bar q$ 
and $q^*\bar{q}^* \to q \bar q$ mechanisms.

\section{Numerical results}
\label{sec:results}

In this section we will present results for the color-singlet
gluon-gluon fusion and color evaporation model.

To convert to the number of $f_0(980)$ mesons per event, 
as was presented in Ref.~\cite{Lee:thesis}, 
we use the following relation:
\begin{equation}
\frac{d N}{d p_t} = \frac{1}{\sigma_{{\rm inel}}} 
                    \frac{d \sigma}{d p_t} \,.
\label{dN_dpt}
\end{equation}
The inelastic cross section for $\sqrt{s} = 7$~TeV 
was measured at the LHC and is:
\begin{eqnarray}
\sigma_{{\rm inel}} &=& 73.15 \pm 1.26 \,{\rm(syst.)}\, {\rm mb}
\,,\\
\sigma_{{\rm inel}} &=& 71.34 \pm 0.36 \,{\rm(stat.)} \pm 0.83 \,{\rm(syst.)} \, {\rm mb} \,,
\label{sigma_ine}
\end{eqnarray}
as obtained by the TOTEM \cite{Antchev:2013gaa} 
and ATLAS \cite{Aad:2014dca} collaborations, respectively. 
In our calculations we take $\sigma_{{\rm inel}} = 72.5$~mb.

\subsection{Gluon-gluon fusion}

As discussed in the previous section the result of the
color-singlet gluon-gluon fusion strongly depends on the flavor
structure of $f_0(980)$ which is related to the $(<e_{q}^2>)^2$ in
Eq.~(\ref{replacement}). For example 
$(<e_{q}^2>)^2 = 25/162$ for first scenario (\ref{nnbar}), 
$(<e_{q}^2>)^2 = 1/81$ for the $s \bar s$ scenario (\ref{ssbar}).
For the tetraquark scenario $(<e_{q}^2>)^2 = 1/162$ (\ref{tetraquark})
assuming diquark as elementary object, 
but everything depends on details and assumptions made for diquark.

\begin{figure}[!ht]
\includegraphics[width=0.495\textwidth]{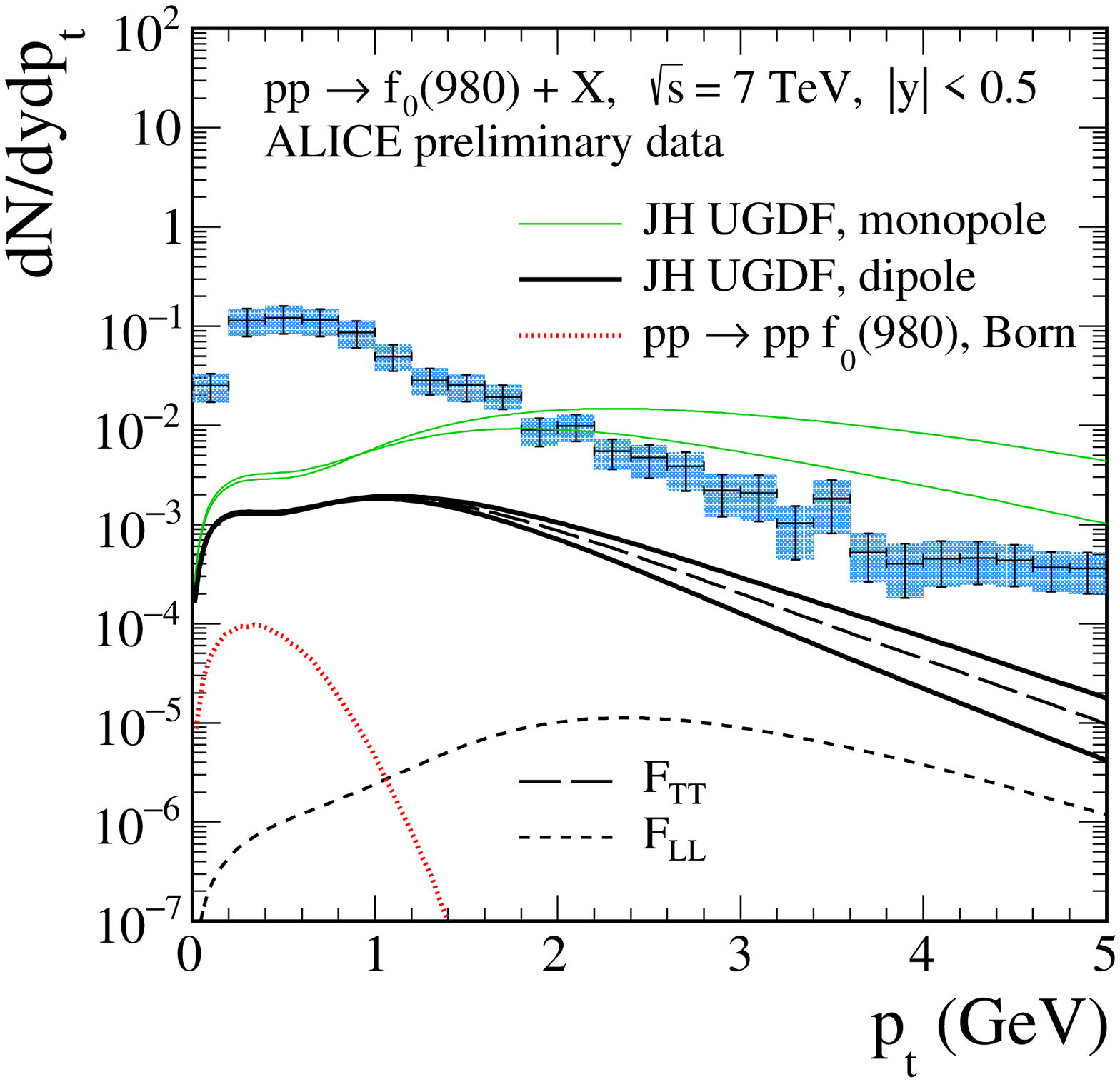}
\includegraphics[width=0.495\textwidth]{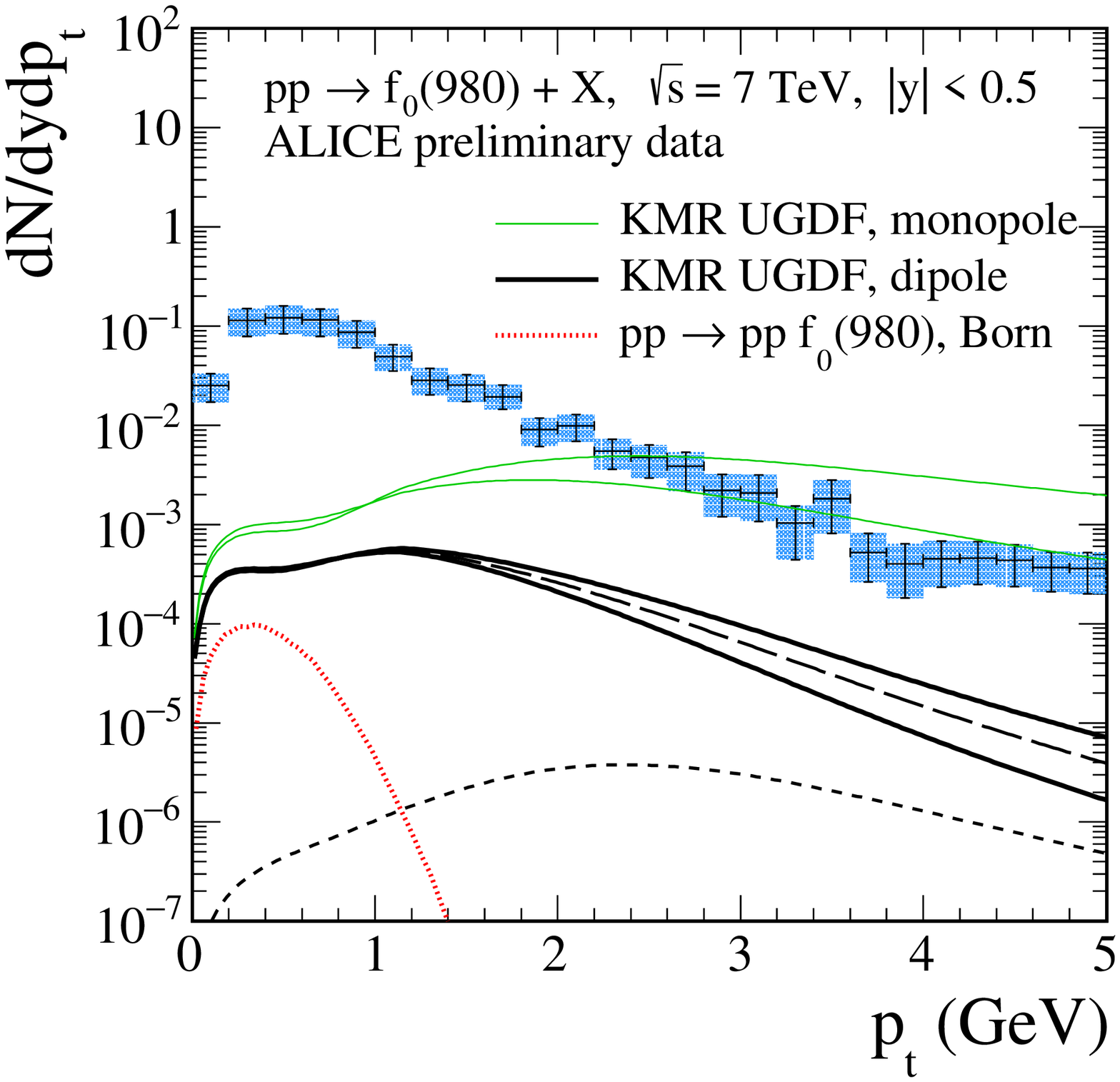}
\caption{\label{fig:dN_dpt_TTLL}
\small
The $f_{0}(980)$ meson transverse momentum distributions at
$\sqrt{s}=7$~TeV and $|{\rm y}|<0.5$.
The preliminary ALICE data from \cite{Lee:thesis} are shown for comparison.
For the $g^* g^* \to f_0(980)$ contribution two different 
UGDFs are used: the JH (left panel) and KMR (right panel). 
Here, the $s \bar{s}$ flavor wave function of $f_0(980)$ 
is taken into account.
Shown are TT and LL components in the amplitude
and their coherent sum (total) 
for the monopole (green solid lines) 
and dipole (black solid lines) form factor parametrizations. 
In this calculation we used $\Lambda_M = \Lambda_D = m_{f_{0}}$
and $R_{LL/TT} = \pm 0.5$.
The upper solid lines are for $R_{LL/TT} = -0.5$ 
while the bottom solid lines for 0.5.
The dotted line corresponds to the Born-level result 
for the $pp \to pp f_0(980)$ process via pomeron-pomeron fusion.}
\end{figure}

In Fig.~\ref{fig:dN_dpt_TTLL} we present 
the $f_{0}(980)$ meson transverse momentum distributions at
$\sqrt{s}=7$~TeV and $|{\rm y}|<0.5$.
Here we show results for the color-singlet gluon-gluon fusion contribution
for the $s \bar{s}$ scenario for two different UGDFs,
JH UGDF (left panel) and KMR UGDF (right panel)
together with the preliminary ALICE data from \cite{Lee:thesis}.
These UGDFs are available from the {\tt CASCADE} Monte Carlo code 
\cite{Jung:2010si}:
\begin{itemize}
\item 
We use a glue constructed according to the prescription
initiated in \cite{Kimber:2001sc} and later updated in 
\cite{Watt:2003vf,Martin:2009ii},
which we label as ``KMR UGDF''.
\item 
The second type of UGD which we use has been obtained 
by Hautmann and Jung \cite{Hautmann:2013tba} 
from a description of precise HERA data 
on deep inelastic structure function 
by a solution of the CCFM evolution equation
\cite{Ciafaloni:1987ur,Catani:1989yc,Catani:1989sg}. 
We use ``JH-2013-set2'' of Ref.~\cite{Hautmann:2013tba},
which we label as ``JH UGDF''.
\end{itemize}
We show results for the monopole~(\ref{monopole}) 
and dipole~(\ref{dipole}) form factors
with the cut-off parameter $\Lambda_M = \Lambda_D = m_{f_{0}}$.
For the LL form factor~(\ref{F_LL_parametrization}) 
we take $R_{LL/TT} = \pm 0.5$.
The upper solid lines are for $R_{LL/TT} = -0.5$ 
while the bottom lines for 0.5.
The JH UGDF (see the left panel) gives slightly
larger cross section than the KMR UGDF (see the right panel).
The theoretical distribution for the monopole form factor
with $\Lambda_M = m_{f_{0}}$ exceeds the ALICE data 
for $p_t > 2$~GeV.

The obtained results are much below the preliminary ALICE 
data \cite{Lee:thesis} at low $f_0(980)$ transverse momenta. 
Does it mean that other mechanism(s) is (are) at the game? 

It seems that even the $s \bar{s}$ scenario does not allow to
describe the ALICE data.
A big gluonic component in the $f_0(980)$
wave function could help to improve the situation. 
Large $K \bar{K}$ molecular component could be another solution. 

In addition to the gluon-gluon fusion contribution
we show the contribution 
of the exclusive $pp \to pp f_0(980)$ process
proceeding via the pomeron-pomeron fusion mechanism.
The result is represented by the red dotted line.
Here the calculation was made in the tensor-pomeron approach 
in the Born approximation (without absorptive corrections).
Absorption corrections are important only when restricting to purely
exclusive processes.
For details regarding this approach we refer to 
\cite{Ewerz:2013kda,Lebiedowicz:2013ika,Lebiedowicz:2016ioh,Lebiedowicz:2018eui}.
In the calculation we take the pomeron-pomeron-$f_0(980)$
($\Pom \Pom f_{0}(980)$) coupling parameters from \cite{Lebiedowicz:2018eui},
that is, $(g',g'') = (0.53, 2.67)$; see Table~II of \cite{Lebiedowicz:2018eui}.
We have checked, that with these parameters we describe, within experimental errors, 
the cross sections reported very recently by 
the CMS Collaboration \cite{CMS:pippim}
for the exclusive $pp \to pp (f_0(980) \to \pi^{+} \pi^{-})$ process.

\begin{figure}[!ht]
\includegraphics[width=0.495\textwidth]{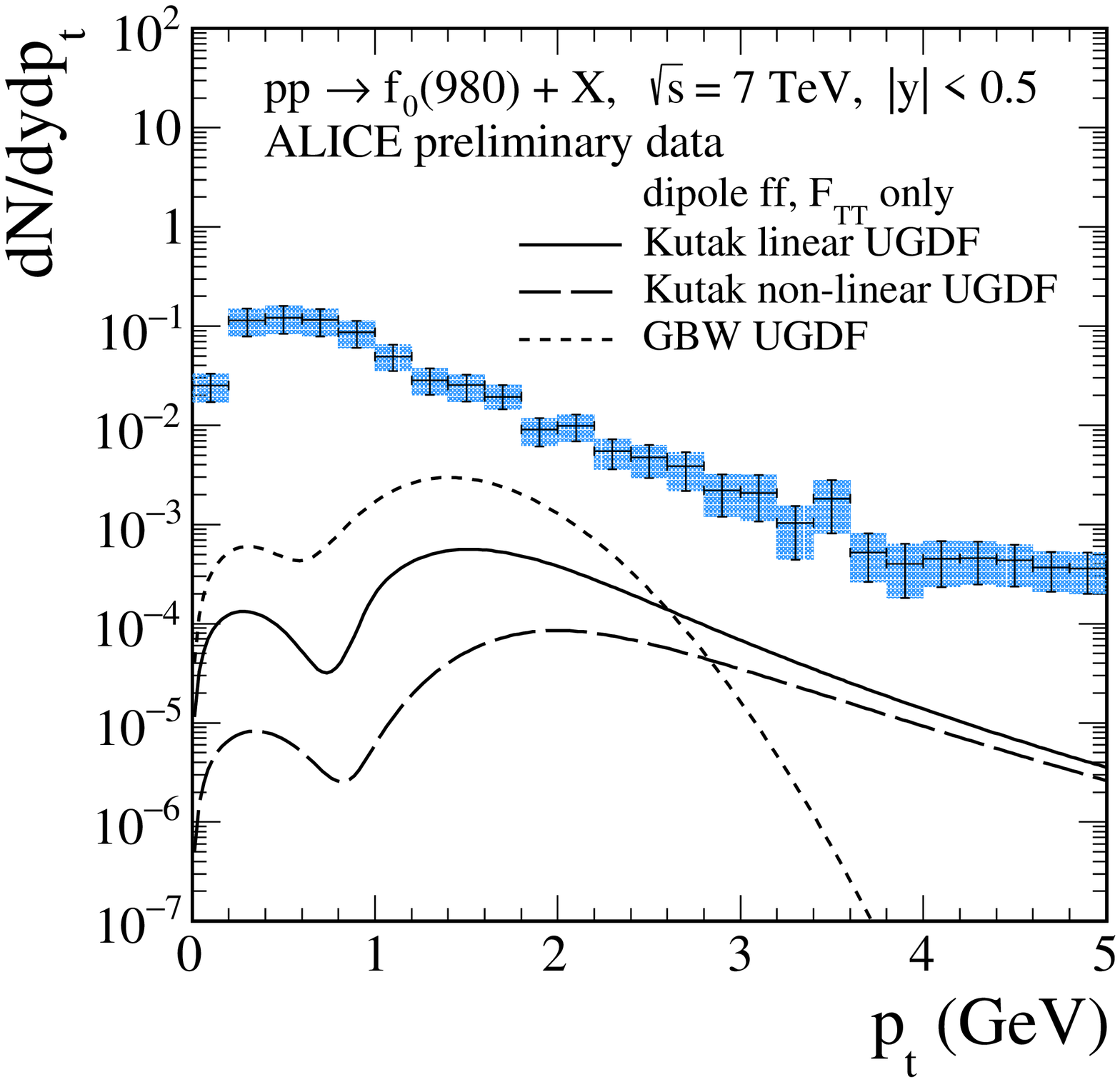}
\includegraphics[width=0.495\textwidth]{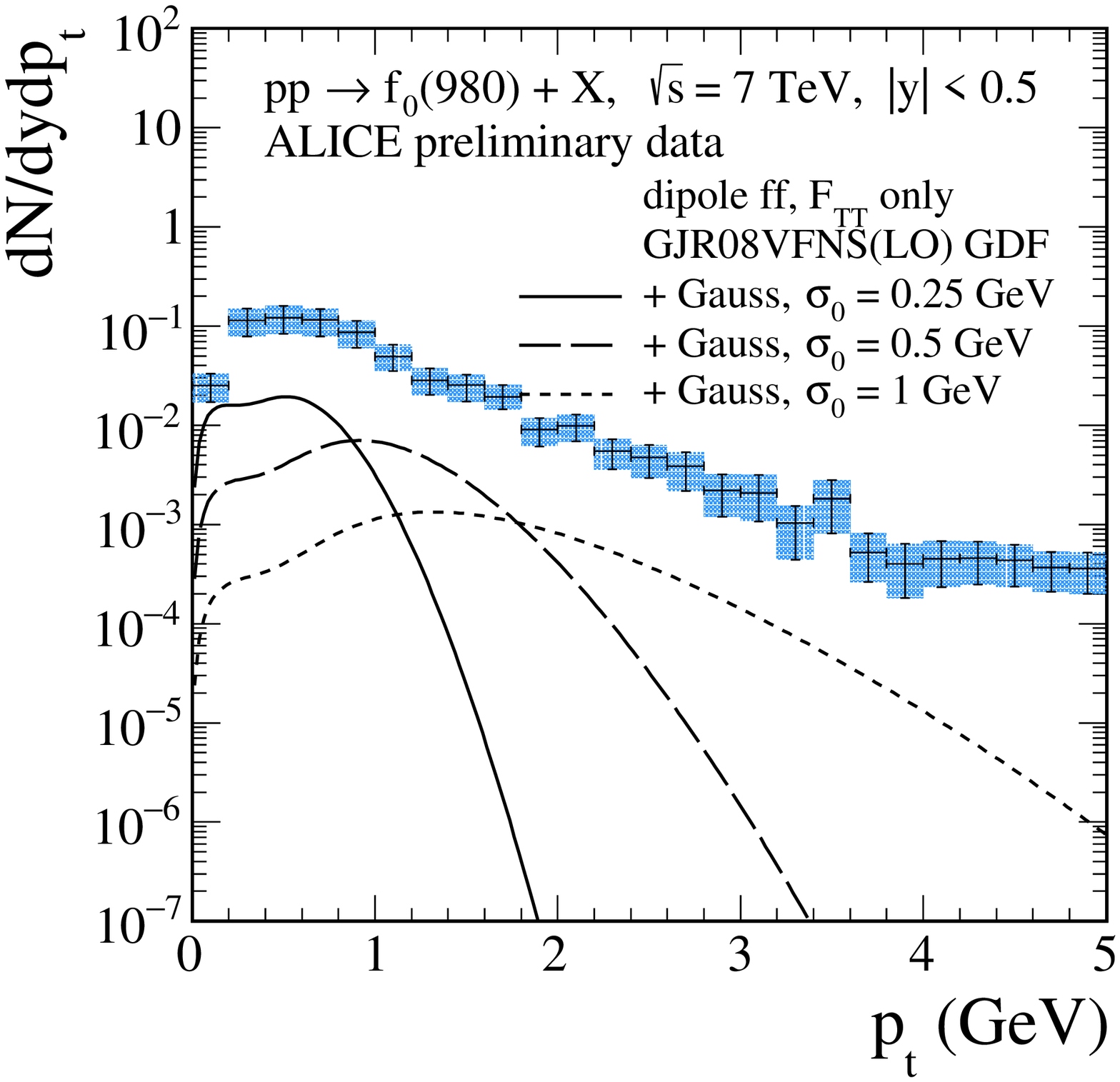}
\caption{\label{fig:dsig_dpt_other_UGDFs}
\small
Transverse momentum distribution of $f_0(980)$ for different
UGDFs from the literature. 
In the left panel we show the results for
Kutak linear (solid line), Kutak non-linear (long-dashed line) 
and GBW (dotted line) UGDFs. 
In the right panel we present our study of the dependence 
on the Gaussian smearing parameter $\sigma_{0}$. 
Here GJR08VFNS(LO) GDF \cite{Gluck:2008gs} was used.}
\end{figure}

In the considered cases (gluon-gluon fusion) 
we observe relatively quick drop of
$d\sigma/dp_t$ for $p_t \to 0$.
Is it a specific feature of the considered UGDFs (JH or KMR)?
In Fig.~\ref{fig:dsig_dpt_other_UGDFs} we show the dominant $TT$
contribution to $d\sigma/dp_t$ for other UGDFs used in the literature. 
In the left panel we show results for the GBW UGDF \cite{GolecBiernat:2001mm} 
(from {\tt CASCADE} code)
and the Kutak UGDF \cite{Kutak:2014wga}.
We have used two versions of the Kutak's UGDF.
Both introduce a hard scale dependence via a Sudakov form factor 
into solutions of a small-$x$ evolution equation.
The first version (linear) uses a BFKL evolution with a resummation
of subleading terms. 
The second version (non-linear) uses an evolution equation of 
Balitsky-Kovchegov type.
The non-linear version leads to smaller cross section,
especially for small $f_{0}(980)$ meson transverse momenta.
To better illustrate the dependence of UGDFs on $\bq^2$ in the right panel 
we present similar results with the Gaussian smearing of collinear GDF, 
often used in the context of TMDs, 
for different smearing parameter $\sigma_{0} = 0.25, 0.5, 1.0$~GeV.
The GJR08VFNS(LO) collinear GDF \cite{Gluck:2008gs} 
was used for this purpose.
As expected the shape of $d\sigma/dp_t$ strongly depends on
the value of the smearing parameter $\sigma_{0}$ used in the calculation.
The speed of approaching $d\sigma/dp_t$ for $p_t \to 0$
strongly depends on the value of $\sigma_{0}$. 
It is impossible to describe simultaneously $p_t < 1$~GeV 
and $p_t > 1$~GeV regions with the same value of $\sigma_{0}$. 
This illustrates the generic situation with all UGDFs.

The $d\sigma/dp_t \to 0$ for $p_t \to 0$ is related to
the behavior of 
${\cal F}_{g}(x,{\bq}^2,(\mu_{F}^2))$ for $\bq^2 \to 0$ because
$\bqa + \bqb = \bp$.
This nonperturbative part is rarely discussed by the UGDF builders and
is often done by an extrapolation from the perturbative region of 
$\bq^2 > (0.5 - 1.0)$~GeV$^2$, trying to satisfy:
\begin{equation}
\int^{\mu_{F}^2} d\bq^2 {\cal F}_{g}(x,\bq^2,(\mu_{F}^2))  
\sim x g(x,\mu_{F}^2)\,,
\label{UGDF_normalization}
\end{equation}
or trying to describe deep-inelastic scattering data, 
especially from HERA.
Above $g(x,\mu^2)$ is the collinear GDF.

\begin{figure}[!ht]
\begin{center}
\includegraphics[width=0.45\textwidth]{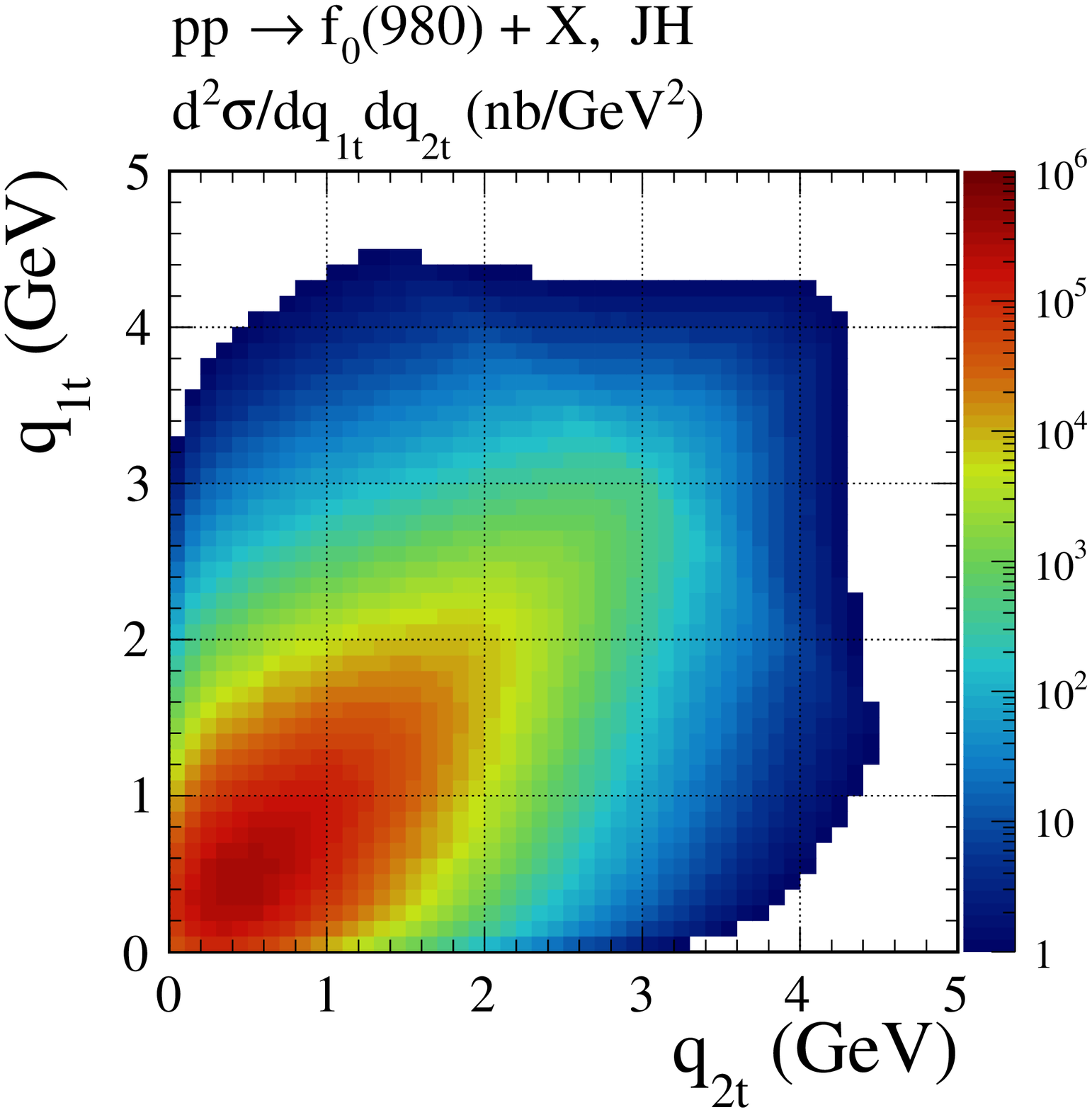}
\includegraphics[width=0.45\textwidth]{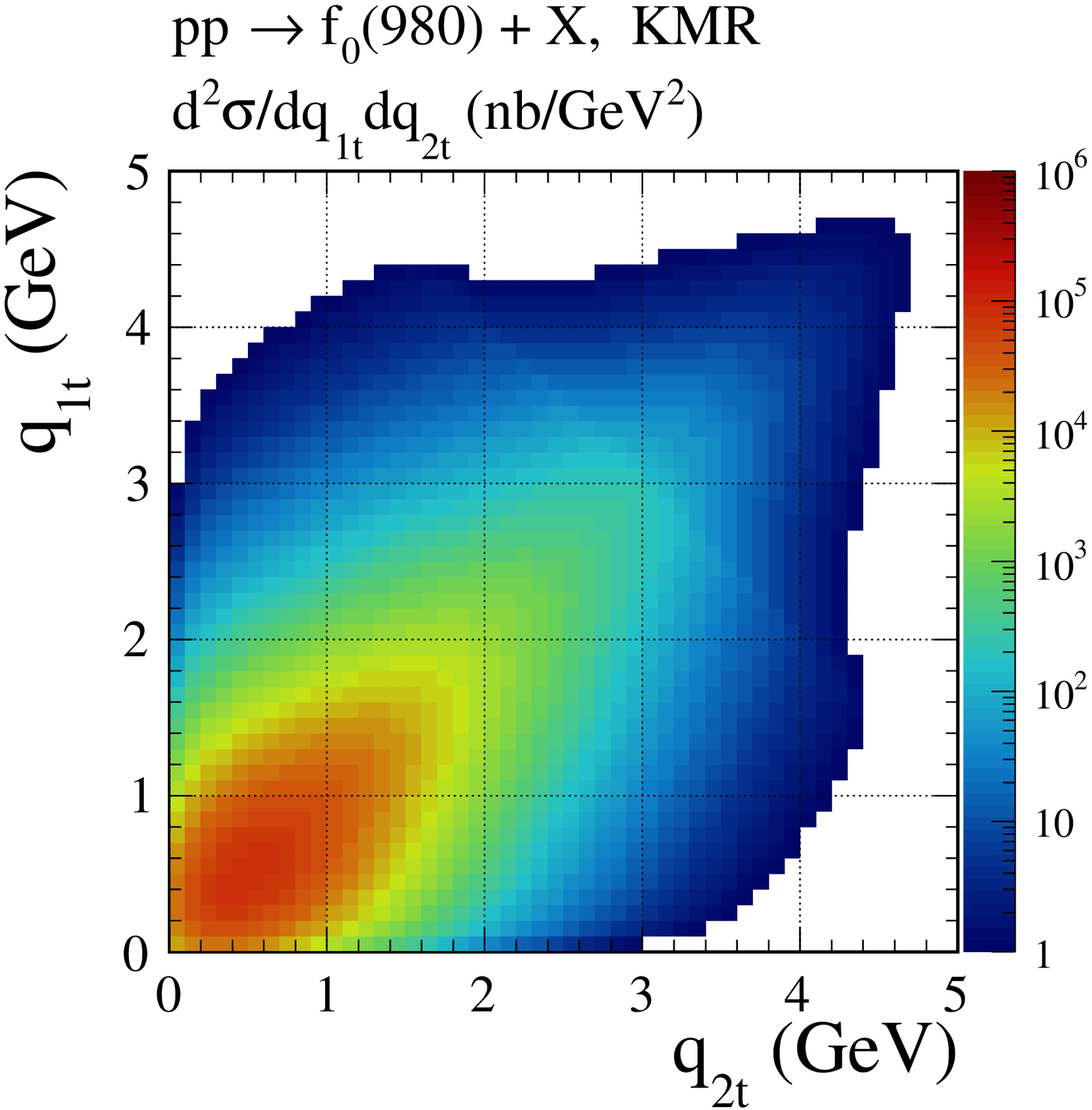}\\
\includegraphics[width=0.45\textwidth]{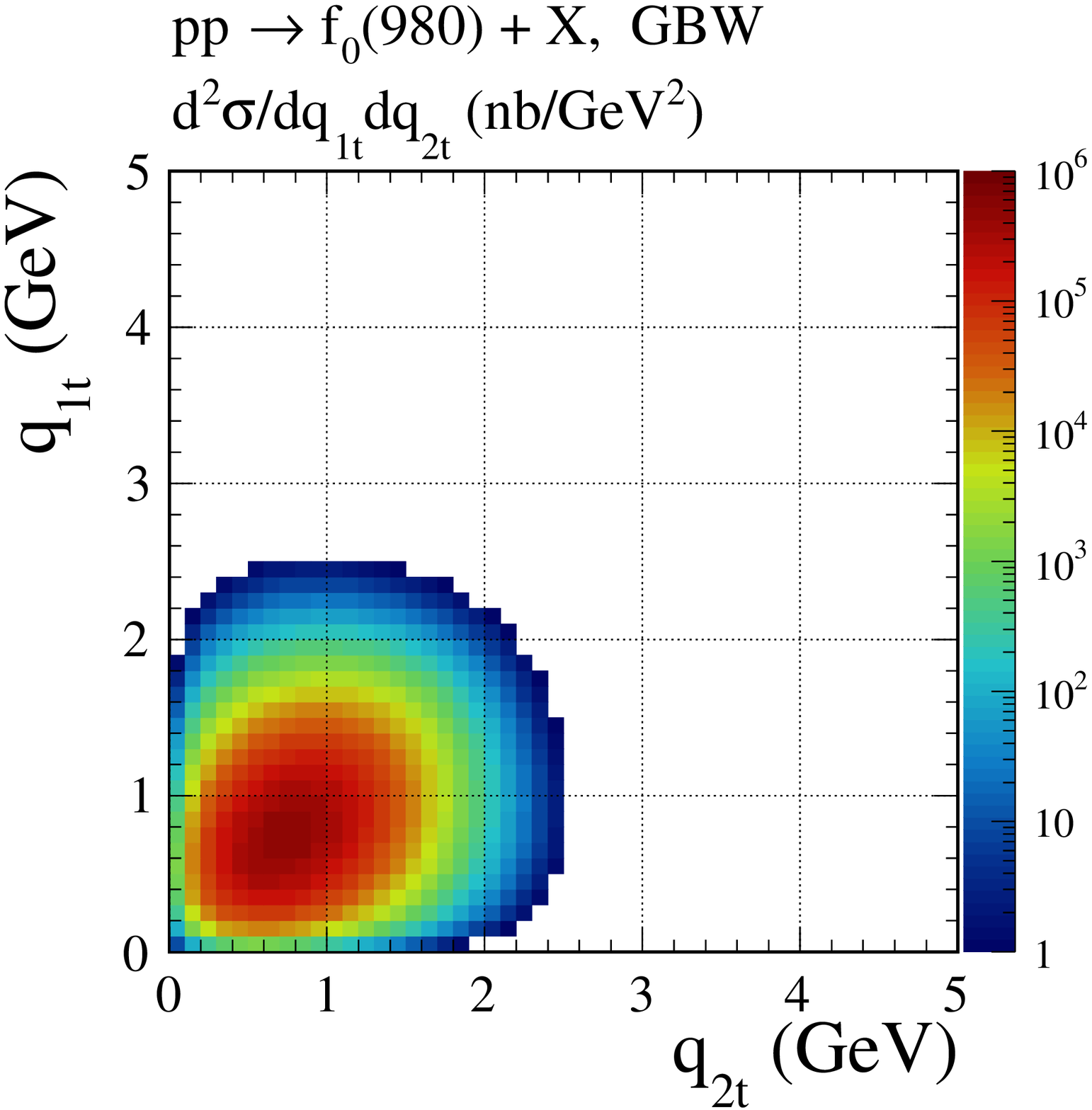}
\includegraphics[width=0.45\textwidth]{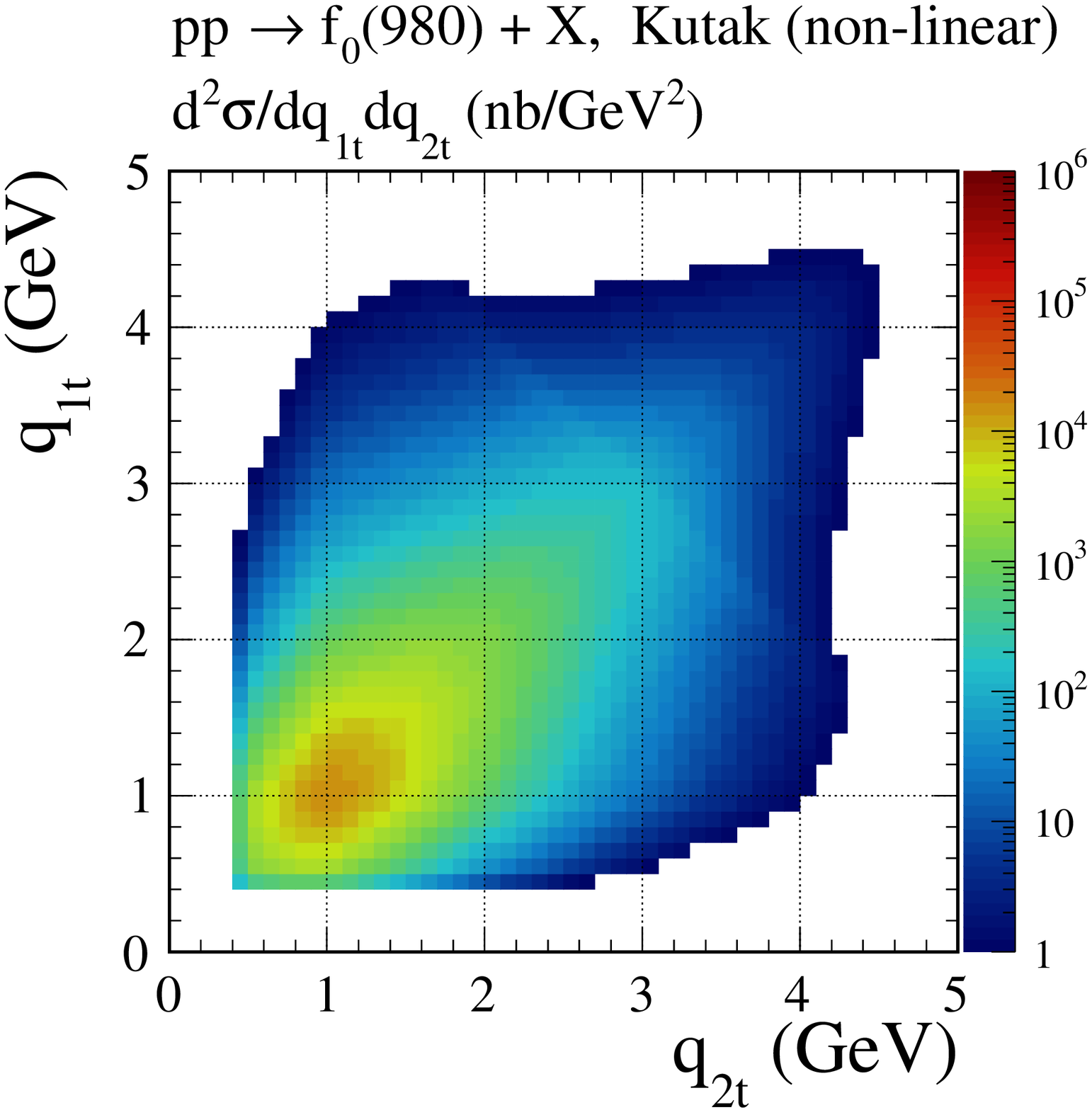}
\end{center}
\caption{\label{fig:dsig_dq1tdq2t}
\small
Two-dimensional distributions in gluon transverse momenta
for different UGDFs from the literature.}
\end{figure}

Such a procedure(s) is (are) of course not unique.
To illustrate how differently it is done for different UGDFs in
Fig.~\ref{fig:dsig_dq1tdq2t} we present a few examples of
$d^{2}\sigma/dq_{1t} dq_{2t}$ for the JH, KMR,
GBW, and Kutak (non-linear) UGDFs.
The maximal contributions, even to the $p_t$-integrated cross section
come from the region of rather small gluon transverse momenta
$q_{1t}, q_{2t} < 1$~GeV.
This is domain of nonperturbative physics which is not fully under
control. The GBW UGDF was destined for this region.
On the other hand the larger-$p_t$ region ($p_t > 2$~GeV) is sensitive
to $q_{1t}, q_{2t} > 1$~GeV where perturbative methods apply.
The fault visible at $q_{1t} + q_{2t}$ = 6 GeV for some distributions
is due to limit of integration over $p_{t}$ to $p_{t} \in (0,6)$~GeV.
The sudden drop of the cross sections for the Kutak UGDF for 
$q_{1t} < 0.4$~GeV or $q_{2t} < 0.4$~GeV is of purely technical nature 
and comes from the limited grid in gluon transverse momentum 
which was available to us. We decided not to make extra efforts to
extrapolate the grids down to zero.

It is too risky, in our opinion, to use the $d\sigma/dp_t$ data
for $f_0(980)$ production to model UGDFs, as it is improbable that
the $g^* g^* \to f_0(980)$ is the only one.
In addition, the result for $d\sigma/dp_t$ strongly
depends on the assumption done for the flavor structure of $f_0(980)$
as discussed above.
It is not excluded that production of other light/heavy mesons for
small $p_{t}$ can be used to constrain ${\cal F}_{g}(x,{\bq}^2,(\mu_{F}^2))$ at small ${\bq}^2$.

Below we shall consider also color octet contribution
calculated in the color evaporation approach.

\subsection{Color evaporation model (CEM)}

In the present study the cross sections for $u\bar u$ and $d\bar d$
or alternatively $s \bar s$
minijet pair production are calculated in the $k_{t}$-factorization
approach or in the collinear approach. In both cases the calculations
are done with the help of the {\tt KaTie} Monte Carlo code 
\cite{vanHameren:2016kkz}. 
Considering production of (soft) minijets a real problem is a
regularization of the cross section at small transverse momenta. Here we
follow the methods adopted for collinear approach in {\tt PYTHIA} 
and multiply the calculated cross section by a somewhat arbitrary 
suppression factor:
\begin{eqnarray}
F_{\mathrm{sup}}(p_t) = \frac{p_t^4}{((p_{t}^{0})^{2} + p_t^2)^2}  \,,
\label{F_sup_1}
\end{eqnarray}
where $p_{t}^{0}$ is a free parameter of the model. 
In the following calculations we take different values of $p_{t}^{0}$, in order to  
show sensitivity of the results to the choice of this parameter.
The parameter goes also into the argument 
of the strong coupling constant 
\mbox{$\alpha_{\rm s}(\mu_{R}^2)=\alpha_{\rm s}((p^{0}_{t})^2+p_{t}^2)$}.

\subsubsection{The $k_{t}$-factorization approach to CEM with the KMR UPDFs}

In the $k_{t}$-factorization approach the non-zero $q\bar q$ pair 
transverse momentum can be generated even at leading-order
when only the $2 \to 2$ three-level partonic processes are taken into account. 
Here we include both the $gg$-fusion and $q\bar q$-annihilation mechanisms.
By applying the KMR UPDFs one effectively includes a part of real higher
order corrections. Large amount of extra hard 
emissions present in this model may lead to large transverse momentum 
of the produced system, 
without any additional emissions at the level of hard matrix elements.

Technically, in the numerical calculations here, 
the suppression factor includes the fact 
that the transverse momenta of outgoing minijets 
are not balanced and it takes the following form:   
\begin{eqnarray}
F^{(2)}_{\mathrm{sup}}(p_{1t}^2,p_{2t}^2) = 
\frac{p_{1t}^2}{(p_{t}^{0})^{2} + p_{1t}^2} \times 
\frac{p_{2t}^2}{(p_{t}^{0})^{2} + p_{2t}^2} \,.
\label{F_sup_2}
\end{eqnarray}
The {\tt KaTie} Monte Carlo generator does not have any problems 
with the generation of the events in the case of the $2\to 2$ processes,
even if there is no additional cut-off on the outgoing minijets 
transverse momenta (thus low-$p_{t}$ cuts are not necessary here). 
The generated events for massless quarks/antiquarks 
are weighted by the suppression factor (\ref{F_sup_2}).

\begin{figure}[!ht]
\includegraphics[width=0.495\textwidth]{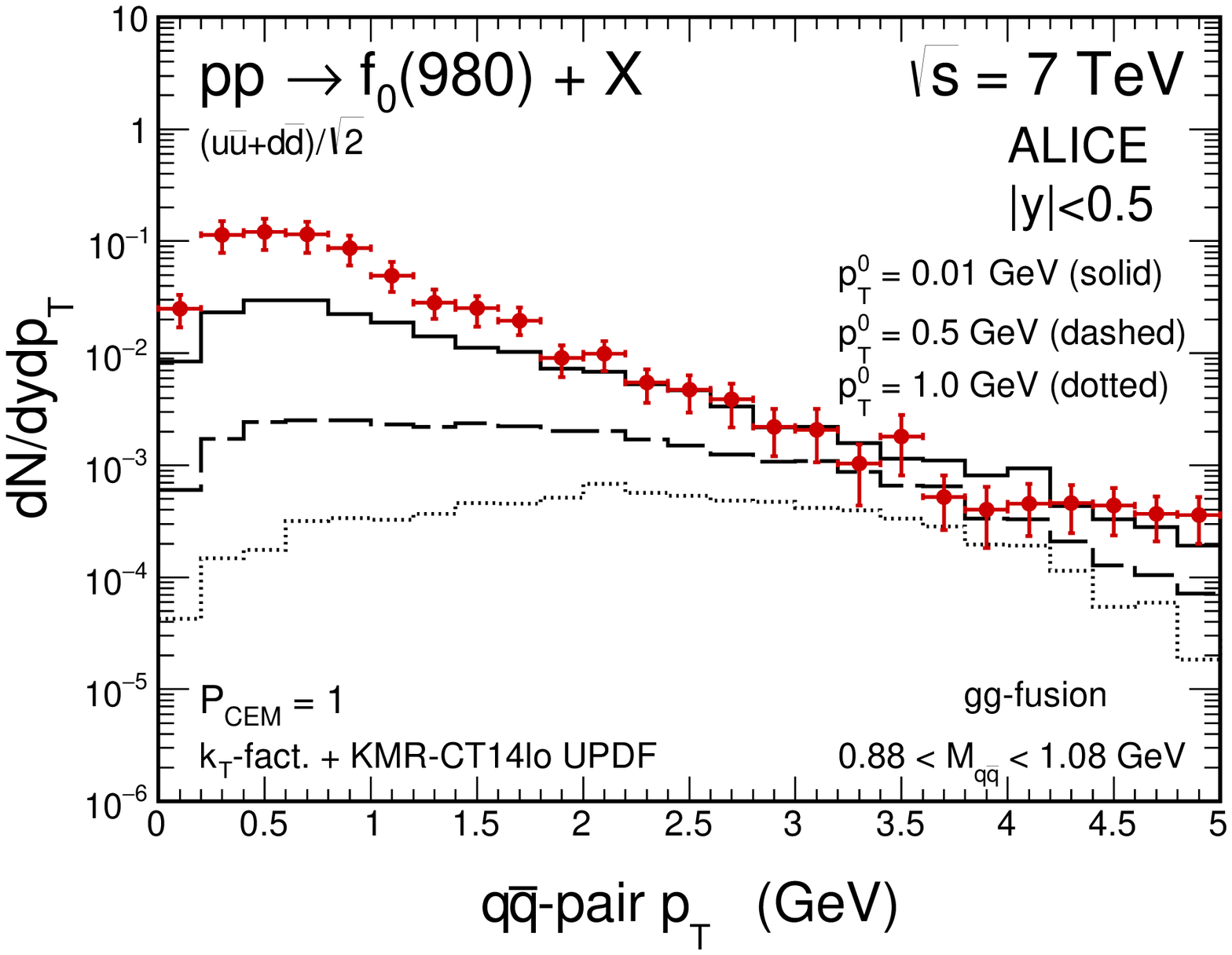}
\includegraphics[width=0.495\textwidth]{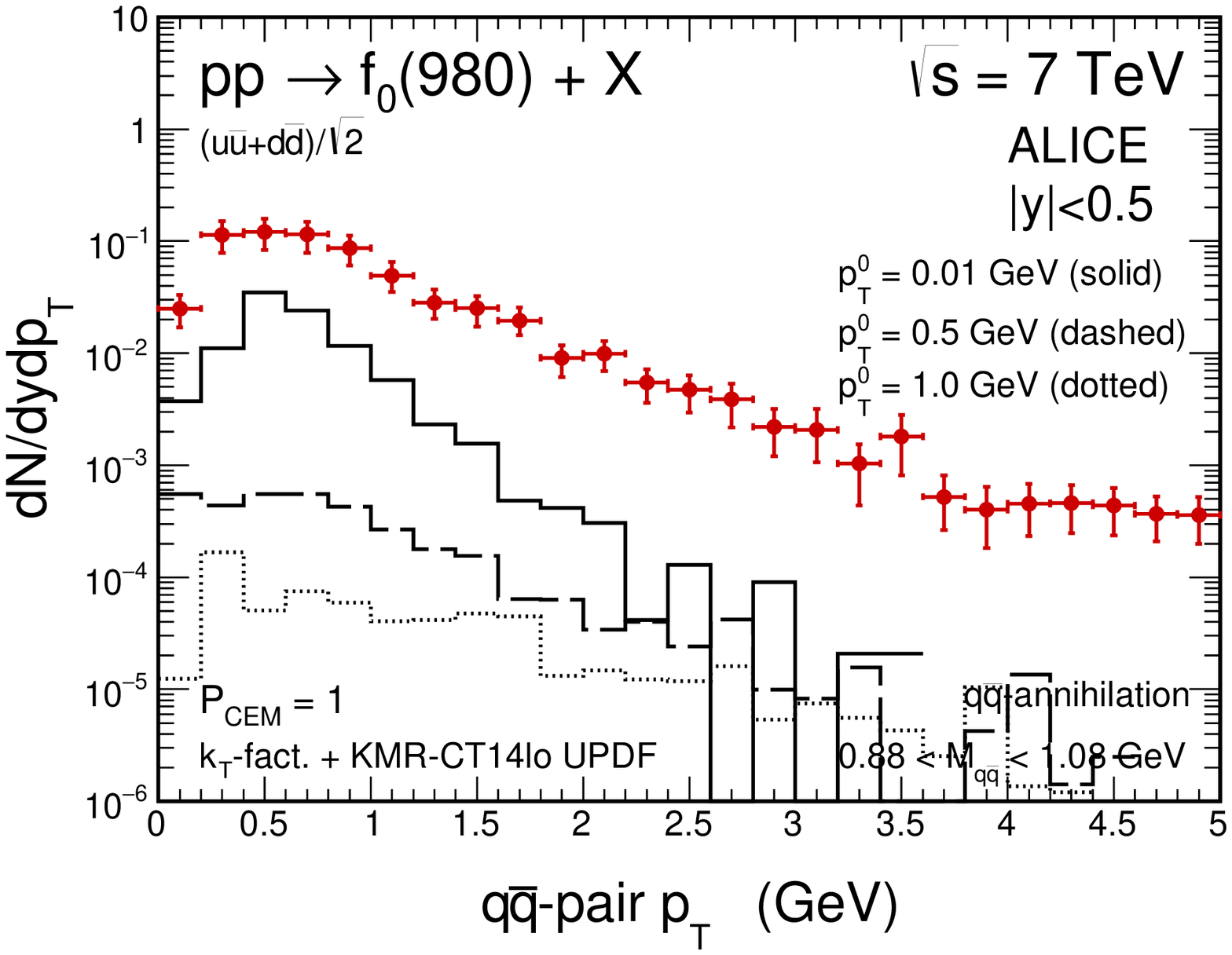}
\caption{\label{fig:KMR_ptcut}
\small
The transverse momentum distribution of $f_0(980)$ for the KMR-CT14lo UPDFs for
different $p_{t}^{0}$ in (\ref{F_sup_2})
for the $gg$-fusion (left) and $q\bar{q}$ (right) mechanisms.
The calculations were done 
for $M_{q \bar{q}} \in (0.88,1.08)$~GeV.}
\end{figure}

In Fig.~\ref{fig:KMR_ptcut} we show the results for different values of $p_{t}^{0}$ in (\ref{F_sup_2}), 
that is,
$p_{t}^{0} = 0.01, 0.5$, and 1.0~GeV.
Large damping of the $q \bar{q}$-pair $p_{t}$ distributions is visible.
In the following, we choose $p_{t}^{0} = 0.01$~GeV.

\begin{figure}[!ht]
\includegraphics[width=0.495\textwidth]{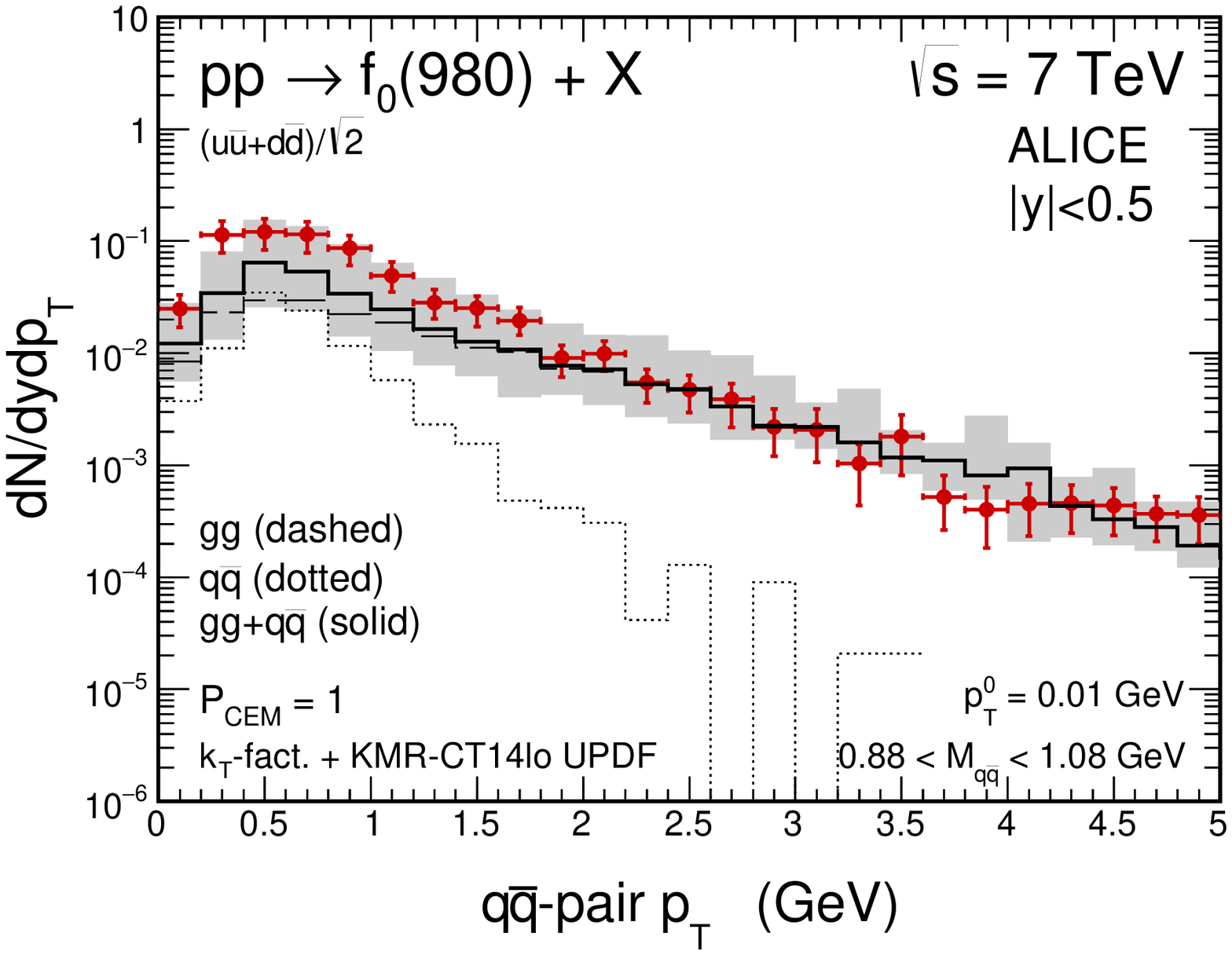}
\includegraphics[width=0.495\textwidth]{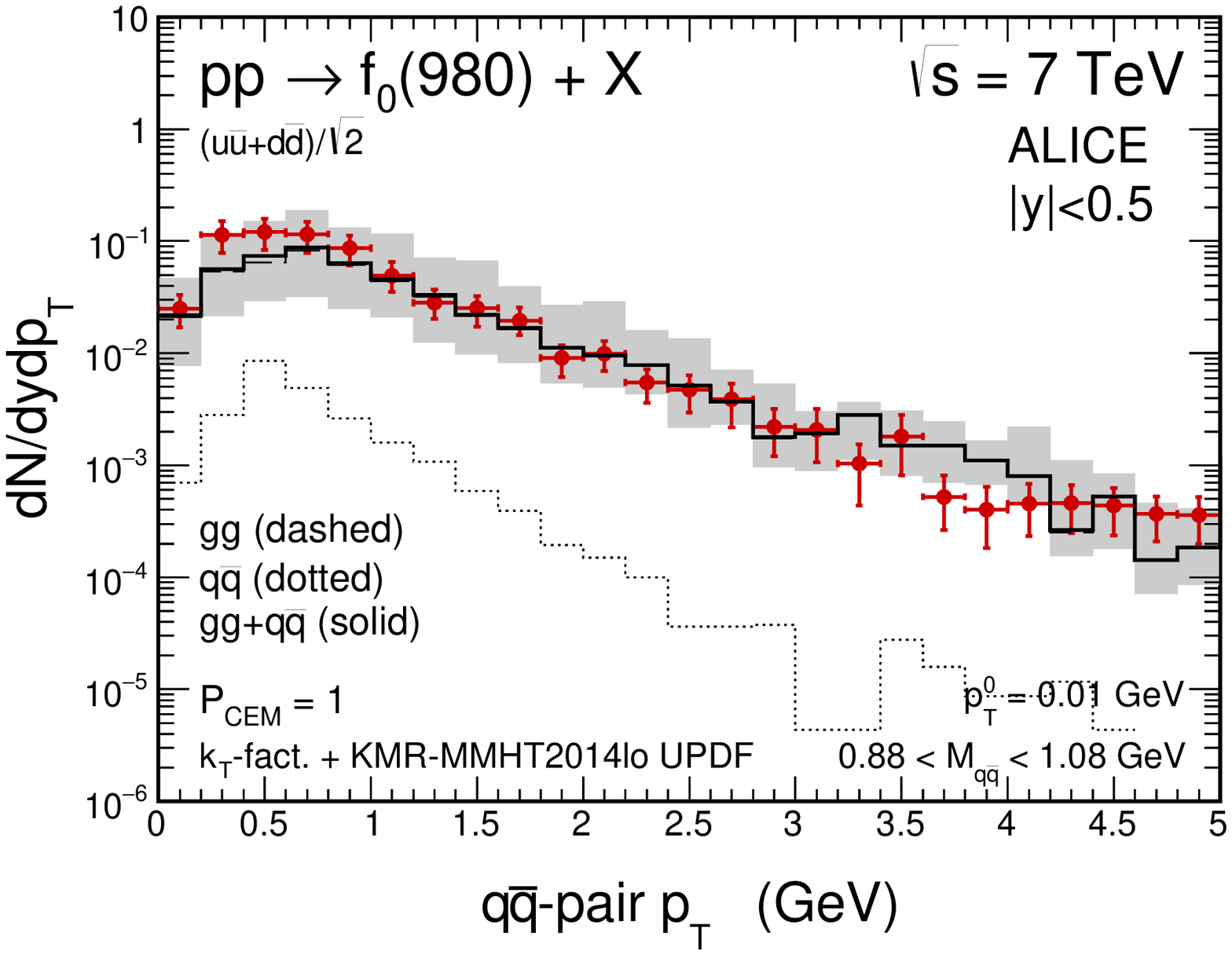}
\includegraphics[width=0.495\textwidth]{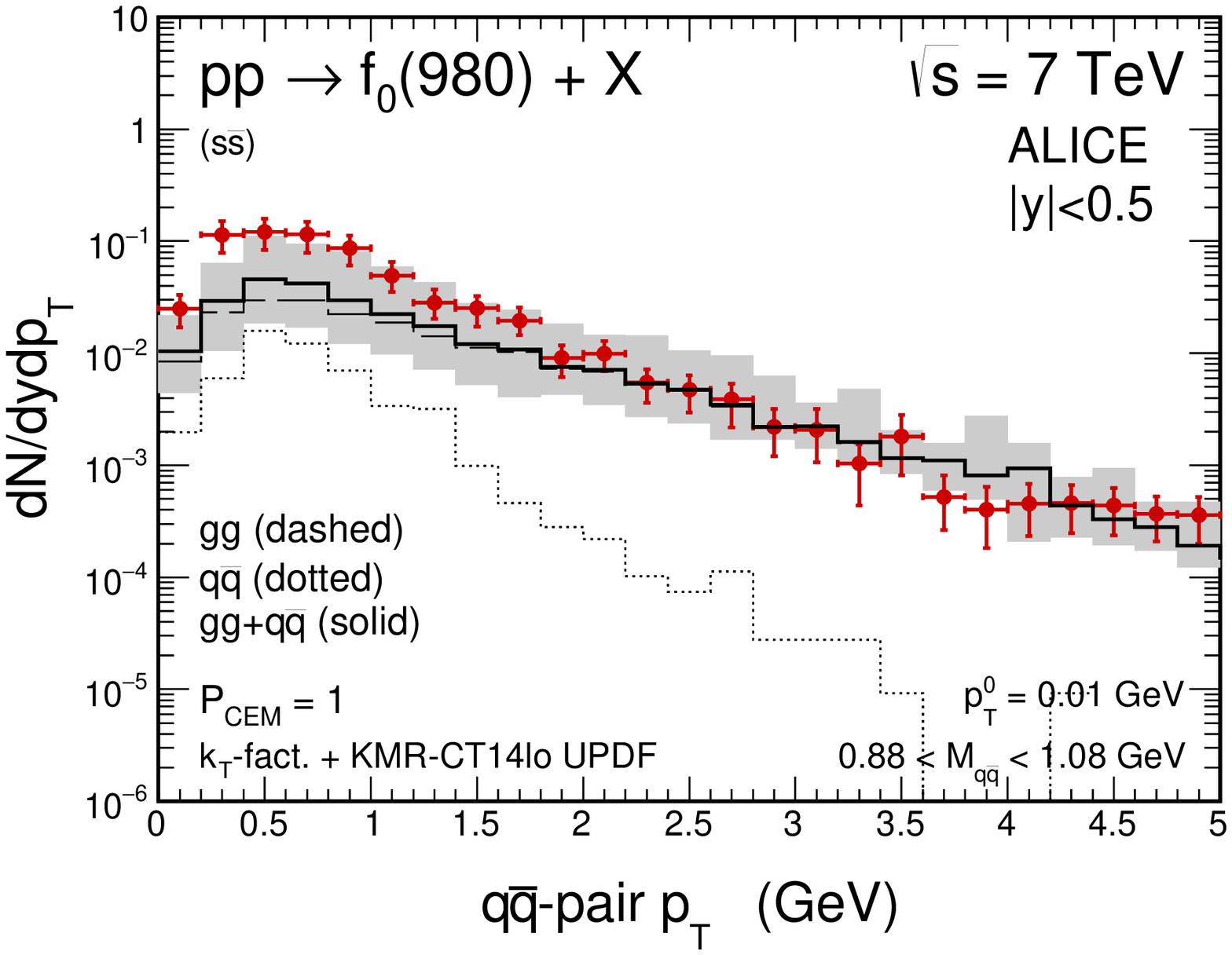}
\includegraphics[width=0.495\textwidth]{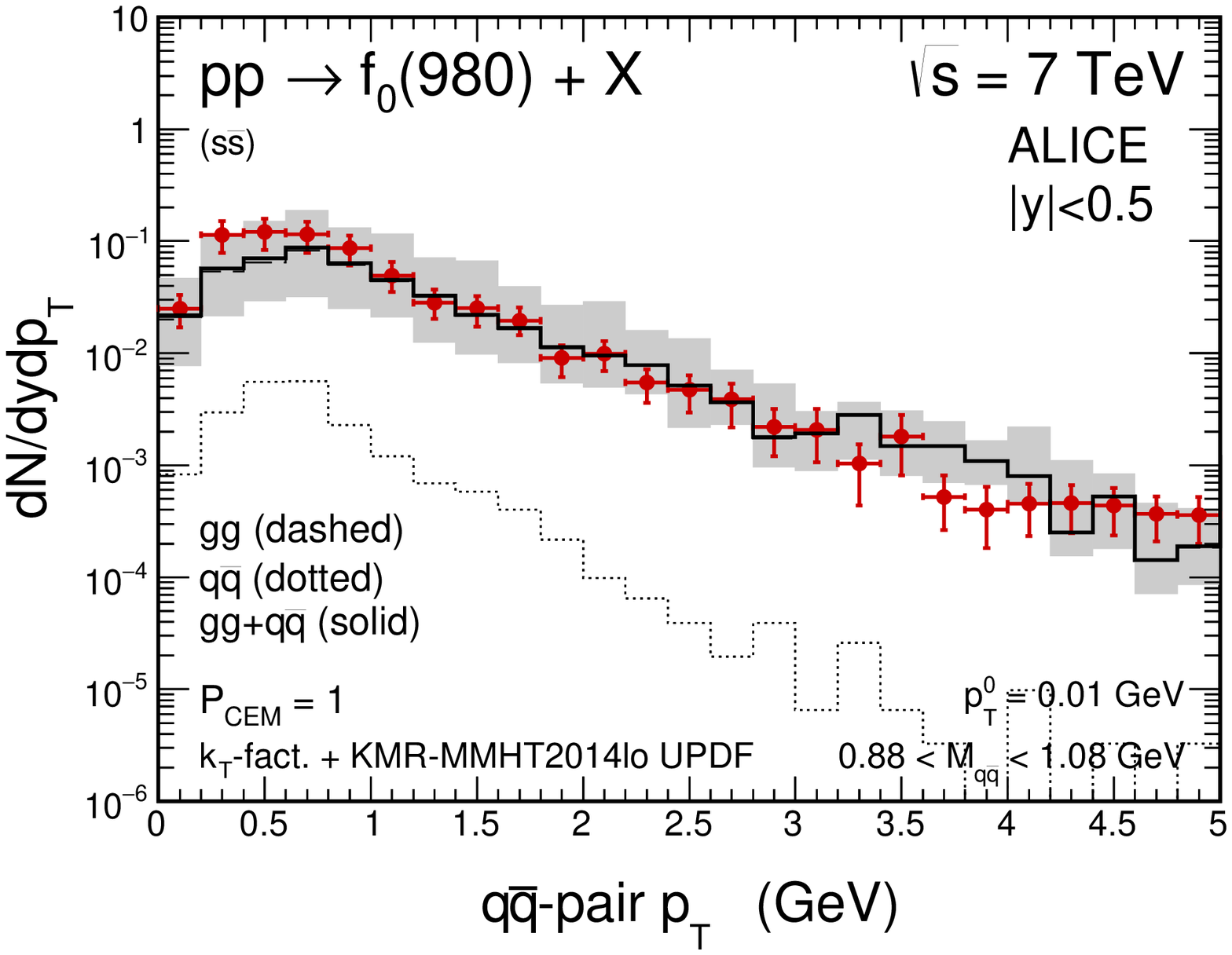}
\caption{\label{fig:KMR}
\small
The $f_{0}(980)$ meson transverse momentum distributions at $\sqrt{s}=7$~TeV 
and $|{\rm y}|<0.5$
calculated in the color evaporation model based 
on the $k_{t}$-factorization approach
using the KMR-CT14lo (left) and KMR-MMHT2014lo (right) UPDFs
together with the preliminary ALICE data from \cite{Lee:thesis}.
The calculations were done in quark-antiquark invariant mass region
$M_{q \bar{q}} \in (0.88,1.08)$~GeV
for the light $q \bar{q}$ scenario (\ref{nnbar}) 
(see the top panels)
and for the $s \bar{s}$ scenario (\ref{ssbar}) (see the bottom panels).
The results for the $gg$-fusion and $q\bar{q}$ mechanisms 
are shown separately.
Their sum is shown by the solid line.
Here we used extremely small $p_{t}^{0} = 0.01$~GeV in (\ref{F_sup_2}).}
\end{figure}

As can be seen from Fig.~\ref{fig:KMR}
we obtain a good description of the ALICE data
even with the leading-order $2 \to 2$ mechanisms only.
In the top panels of Fig.~\ref{fig:KMR} we show results 
for the first $\frac{1}{\sqrt{2}} 
\left( \ket{u \bar{u}} + \ket{d \bar{d}} \right)$ 
scenario (\ref{nnbar})
while in the bottom panels of Fig.~\ref{fig:KMR} 
for the $\ket{s \bar{s}}$ scenario (\ref{ssbar}).
We show also the dependence of the final results 
on the collinear parton distributions 
used to the calculation of the KMR UPDFs.
The results in the left panels correspond 
to the CT14lo PDF \cite{Dulat:2015mca}
while in the right panels to the MMHT2014lo PDF \cite{Harland-Lang:2014zoa}.
The differences at so small scales between different collinear PDFs
could be significant.
Here and in the following the shaded bands represents 
uncertainties of the calculations related to renormalization scale
chosen as an argument in strong coupling $\alpha_{\rm{s}}$. We vary 
the $\mu_{R}$ over the central value, which is set to be averaged 
transverse mass of outgoing particles, by factor 2.

Since the assumption of 
$\frac{1}{\sqrt{2}}( u \bar{u} + d \bar{d})$ for the flavor wave
function of $f_0(980)$ may be not realistic we consider also the 
$s \bar s$ scenario as was done for color-singlet gluon-gluon fusion.
It is obvious that the corresponding cross section will be smaller than that
for the light $q \bar{q}$ scenario. 
In the right bottom panel of Fig.~\ref{fig:KMR} we show
corresponding results for the $s \bar{s}$ scenario. 
It is obvious that here (KMR-MMHT2014lo UPDF) the
$g^* g^* \to s \bar{s} \to f_0(980)$  
is the dominant mechanism.
Assuming massless $s$ and $\bar{s}$ the corresponding cross section is
very similar as for $\frac{1}{\sqrt{2}}( u \bar{u} + d \bar{d} )$
scenario because for high-energy collisions the gluon-gluon fusion is 
the dominant contribution.

\begin{figure}[!ht]
\begin{center}
\includegraphics[width=0.59\textwidth]{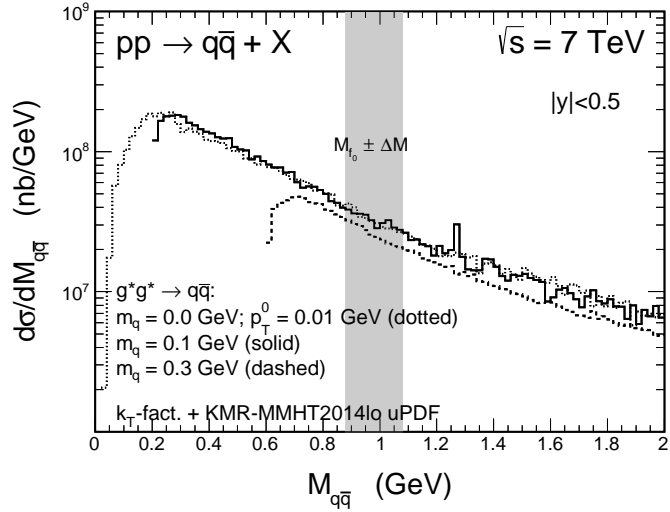}
\caption{\label{fig:invariant mass}
\small
$M_{q \bar q}$ invariant mass distribution for three different
quark/antiquark masses specified in the legend of the figure.}
\end{center}
\end{figure}

The calculations done so far were performed for massless quarks/antiquarks.
How important is the quark/antiquark mass for our $k_t$-factorization
results is illustrated in Fig.~\ref{fig:invariant mass}.
Here we show $M_{q \bar{q}}$ invariant mass distributions for three
different quark masses: $m_q = 0$~GeV 
(with the extra regularization procedure given by Eq.~(\ref{F_sup_2}), $p_{t}^{0} = 0.01$~GeV), 
$m_q = 0.1$~GeV (current $s/\bar s$ mass),
$m_q = 0.3$~GeV (constituent light quark ($u$, $d$) masses).
We show also the window of $M_{q \bar{q}}$ selected for the $f_0(980)$ meson, 
used in the color-evaporation model calculations;
see Eq.~(\ref{transition}).
For finite quark/antiquark masses no extra regularization is needed.
There is no strong dependence on $m_{q}$ provided it is not too big.
For instance for $m_{q} \approx 0.5$~GeV, the $s$ quark constituent mass, 
the cross section for the color evaporation model vanishes when 
$M_{q \bar{q}} > m_{f_0}$.

\begin{figure}[!ht]
\includegraphics[width=0.495\textwidth]{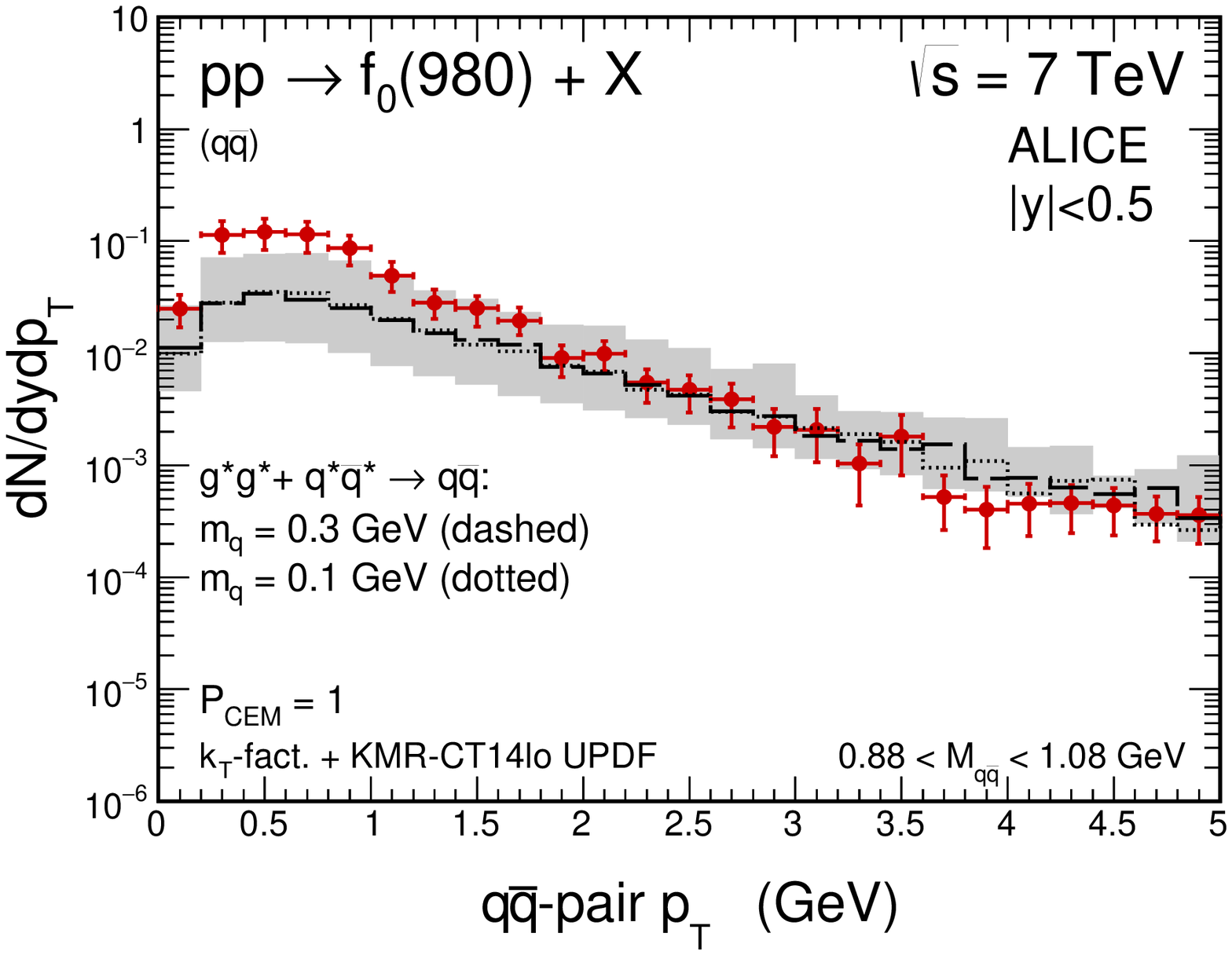}
\includegraphics[width=0.495\textwidth]{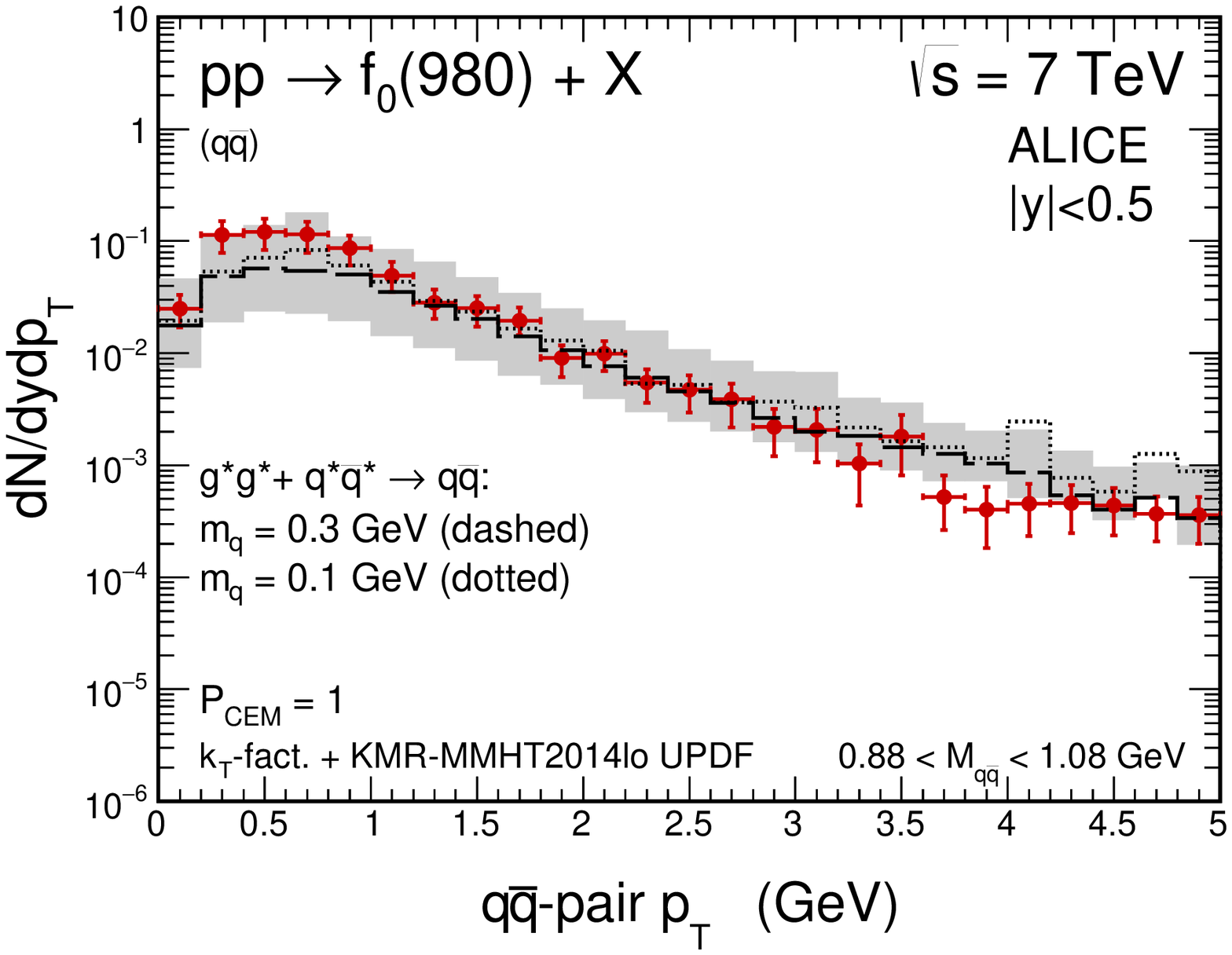}
\caption{\label{fig:KMR_mass}
\small
The transverse momentum distributions of $f_0(980)$ for the KMR UPDFs for two masses of produced quark/antiquark:
$m_{q} = 0.1$~GeV (dotted) and
$m_{q} = 0.3$~GeV (dashed).
The calculations were done in the $q\bar{q}$ invariant mass region
$M_{q \bar{q}} \in (0.88,1.08)$~GeV.}
\end{figure}

In Fig.~\ref{fig:KMR_mass} we show transverse momentum
distribution for the $g^* g^* \to q \bar q$ and $q^* \bar{q}^* \to q
\bar q$ mechanisms added together for different final state
quark/antiquark masses: 0.1, 0.3 GeV. Technically, we use
here off-shell matrix elements derived for heavy quark production
including both the $gg$-fusion and light quark $q\bar q$-annihilation
into heavy (massive) quark-antiquark pair.
We conclude that the results do not depend too much on the mass of
produced quark/antiquark.

\subsubsection{The collinear approach to CEM with the $2 \to 3$ tree-level partonic processes}

In the collinear approach the non-zero $q\bar q$ pair transverse
momentum can be generated only beyond the leading-order 
approximation.
In the calculations we take into account the $2\to 3$ partonic 
processes at the tree-level. So here the $q\bar q$-pair is associated
with extra gluon or quark which comes from the hard matrix elements. 
Here we include all the partonic subprocesses with $gg$-, $qg$- and 
$q\bar q$-types of initial states. 
The full list of the processes
included is shown below:
\begin{itemize}
\item $gg$-fusion:\\ $gg \to g u\bar u$, $gg \to g d\bar d$
\item $qg$-interaction:\\ $gu \to u u\bar u$, $gd \to d u\bar u$, $gs \to s u\bar u$, $g\bar u \to \bar u u\bar u$, $g\bar d \to \bar d u\bar u$, $g\bar s \to \bar s u\bar u$, $ug \to u u\bar u$, $dg \to d u\bar u$, $sg \to s u\bar u$, $\bar u g \to \bar u u\bar u$, $\bar d g \to \bar d u\bar u$, $\bar s g \to \bar s u\bar u$, $gu \to u d\bar d$, $gd \to d d\bar d$, $gs \to s d\bar d$, $g\bar u \to \bar u d\bar d$, $g\bar d \to \bar d d\bar d$, $g\bar s \to \bar s d\bar d$, $ug \to u d\bar d$, $dg \to d d\bar d$, $sg \to s d\bar d$, $\bar u g \to \bar u d\bar d$, $\bar d g \to \bar d d\bar d$, $\bar s g \to \bar s d\bar d$
\item $q\bar q$-annihilation:\\ $u \bar u \to g u\bar u$, $d \bar d \to g u\bar u$, $s \bar s \to g u\bar u$, $ \bar u u \to g u\bar u$, $ \bar d d \to g u\bar u$, $\bar s s \to g u\bar u$, $d \bar d \to g d\bar d$, $u \bar u \to g d\bar d$, $s \bar s \to g d\bar d$, $\bar d d \to g d\bar d$, $\bar u u \to g d\bar d$, $\bar s s \to g d\bar d$
\end{itemize}

\begin{figure}[!ht]
\includegraphics[width=0.495\textwidth]{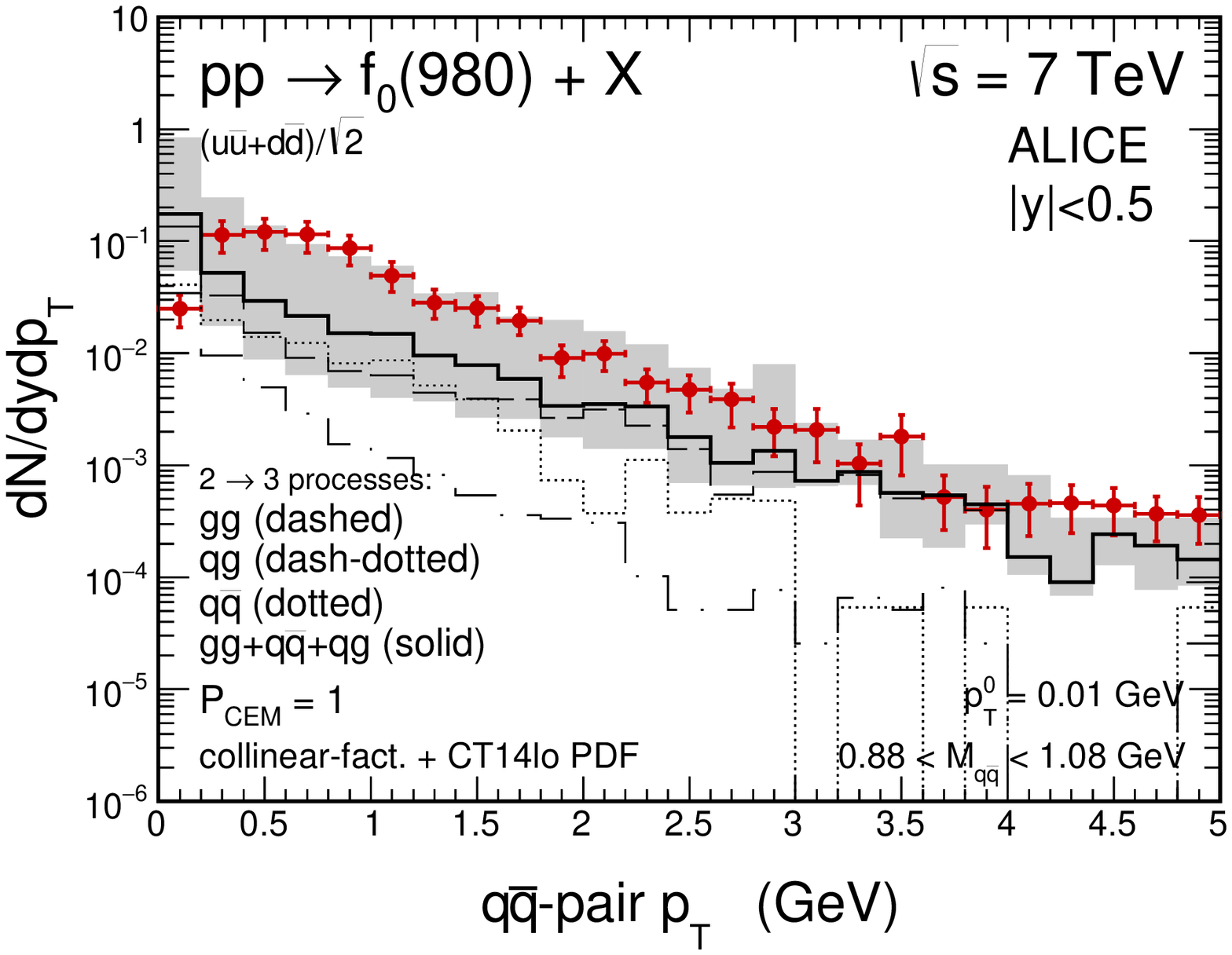}
\includegraphics[width=0.495\textwidth]{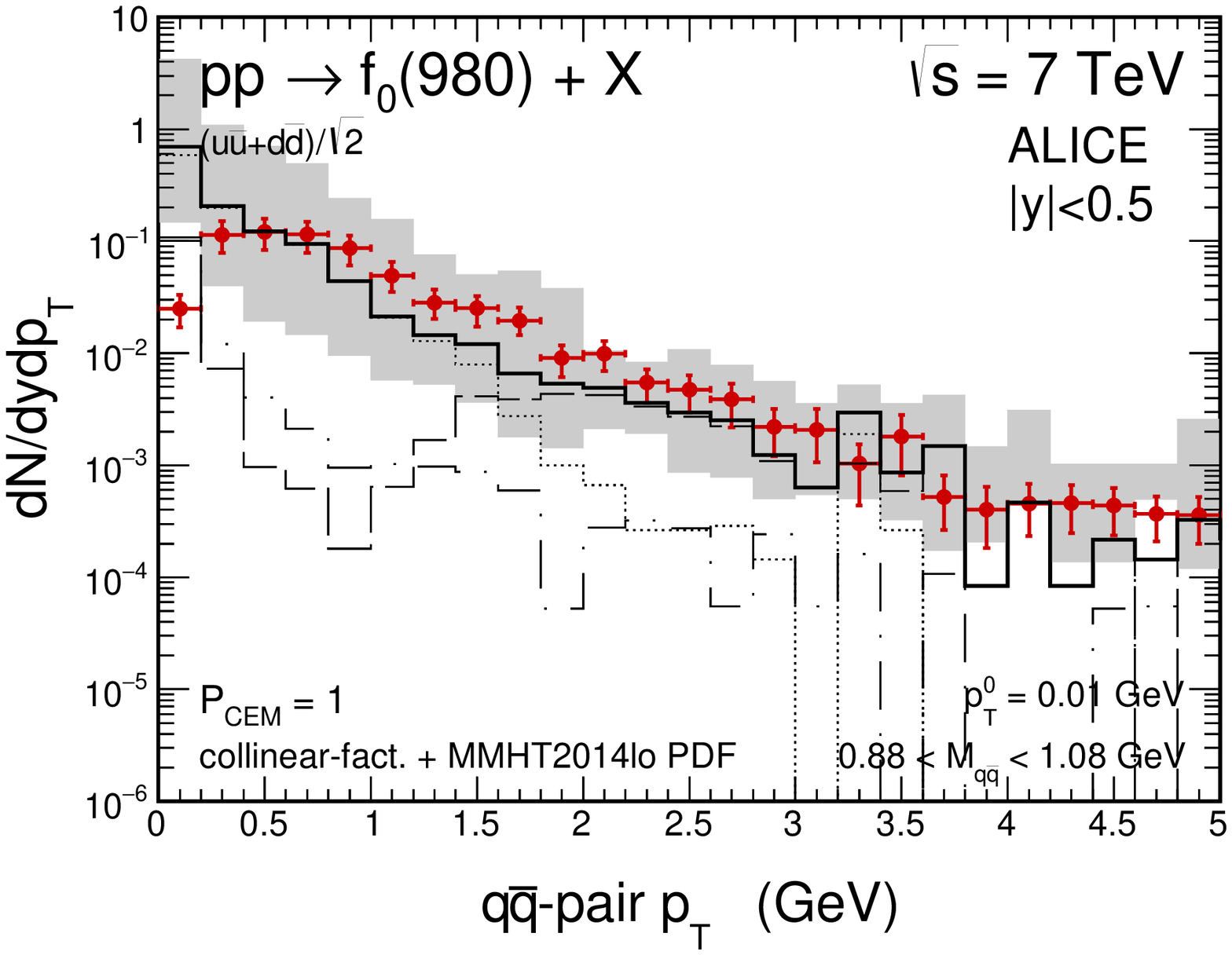}
\caption{\label{fig:coll}
\small
The $f_{0}(980)$ meson transverse momentum distributions at $\sqrt{s}=7$~TeV
and $|{\rm y}|<0.5$, calculated in the color evaporation model based on the collinear approach,
using the CT14lo (left) and MMHT2014lo (right) PDFs together 
with the preliminary ALICE data from \cite{Lee:thesis}.
The calculations were done in quark-antiquark invariant mass region
$M_{q \bar{q}} \in (0.88,1.08)$~GeV.
Here the $gg$, $qg$ and $q\bar q$ induced interaction mechanisms 
are shown separately.
Shown are results for the light $q \bar{q}$ scenario (\ref{nnbar}) for the flavor wave function of $f_0(980)$.
In the calculations we used $p_{t}^{0} = 0.01$~GeV 
in (\ref{F_sup_3}).}
\end{figure}

In the case of the collinear calculations of the $2 \to 3$ processes 
the suppression factor takes the following form:
\begin{eqnarray}
F^{(3)}_{\mathrm{sup}}(p_{1t}^2,p_{2t}^2,p_{3t}^2) = 
\frac{p_{1t}^2}{(p_{t}^{0})^{2} + p_{1t}^2} \times 
\frac{p_{2t}^2}{(p_{t}^{0})^{2} + p_{2t}^2} \times 
\frac{p_{3t}^2}{(p_{t}^{0})^{2} + p_{3t}^2} \,.
\label{F_sup_3}
\end{eqnarray}
The final results that correspond to the
collinear approach are shown in Fig.~\ref{fig:coll}. Again, here we need
to check sensitivity of the results related to the choice of the
collinear PDFs.
In the left panel we show results for the CT14lo PDF while 
in the right panel for the MMHT2014lo PDF.

Clearly, quite different behavior of the calculated 
distributions at small transverse momenta is obtained when comparing the 
results of the $k_t$-factorization and leading-order collinear approach 
for the pair transverse momentum distribution. In the
$k_t$-factorization approach the cross section at very small transverse 
momenta goes down to zero, which is not the case of the collinear 
calculations. This difference has a purely kinematical origin and 
appears as a result of exact treatment of the kinematics with no 
approximations in the $k_{t}$-factorization approach. Similar effect 
was discussed e.g. in Ref.~\cite{Kutak:2016mik} in the case of dijet 
production.

\section{Conclusions}

In this letter we have presented a first exploratory calculation of inclusive
$f_0(980)$ meson production at the LHC energies. Two different mechanisms
have been considered. The first mechanism is the color-singlet 
gluon-gluon fusion known to give a rather good description 
of the $\eta_c$ and $\chi_c$ production 
\cite{Babiarz:2019mag,Babiarz:2020jkh}. 
The second is the color evaporation model 
used e.g. to describe the production of $J/\psi$ meson \cite{Maciula:2018bex}. 
The results have been compared to preliminary ALICE data \cite{Lee:thesis}.

We have started our analysis by considering the 
$\gamma^* \gamma^* \to f_0(980)$ coupling.
Unlike for charmonia we have taken a more phenomenological approach.
The general structure of the $\gamma^* \gamma^* \to f_{0}$ and
$g^* g^* \to f_{0}$ vertices were known from the literature.
However, the corresponding form factors for $g^* g^* \to f_0(980)$ are 
rather poorly known.
The $F_{TT}(0,0)$ has been fixed based on the formula for 
$\Gamma(f_0(980) \to \gamma \gamma)$; see Eq.~(\ref{radiative_decay_width}). 
$F_{LL}(Q_1^2,Q_2^2)$ for $f_0(980)$ is rather unknown and in principle
a model of the $f_0(980)$ wave function is needed.
In the present analysis we have parametrized the $F_{LL}(Q_1^2,Q_2^2)$
form factor in analogy to the results obtained recently
from a microscopic calculation for $\chi_{c0}$ \cite{Babiarz:2020jkh}.
The parametrizations for $F_{TT/LL}(Q_1^2,Q_2^2)$ were restricted only to 
some extent by the Belle data for the $e^+ e^- \to e^+ e^- \pi \pi$ reactions
\cite{Mori:2006jj,Uehara:2008ep}.

Then the $g^* g^* \to f_0(980)$ coupling has been obtained by replacing 
electromagnetic coupling constant by strong coupling constant and by
modifying relevant color factors.

The contribution of color-singlet gluon-gluon fusion strongly depends on
the assumed flavor structure of the $f_0(980)$ meson. For instance
result for the $s \bar{s}$ scenario (\ref{ssbar}) is almost an order of magnitude
larger than that for the light $q \bar{q}$ scenario (\ref{nnbar}).
Large gluonic component in the $f_0(980)$ meson would further increase the
cross section for color-flavor component.

The results for hadroproduction depend on $g^* g^* \to f_0(980)$ form factors
$F_{TT}$ and $F_{LL}$ that have been
parametrized in the present paper; 
see Eqs.~(\ref{monopole})--(\ref{F_LL_parametrization}).
With a plausible parametrization one 
can almost understand transverse momentum distribution of $f_0(980)$ at $p_t >$ 3 GeV
in the $s \bar s$ scenario, but the results for the light
quark/antiquark scenario is much below the data for $p_t <$ 2 GeV.
Clearly a different mechanism is needed to describe the region of
small transverse momenta of $f_0(980)$.
The light $q \bar q$ scenario gives result much below the ALICE data.

In the present paper we have considered also color evaporation mechanism.
Also the color evaporation cross sections have been calculated in 
the $k_t$-factorization approach, as done recently for $J/\psi$
production. The KMR unintegrated parton distribution functions 
(both for gluons, quarks, and antiquarks) have been used in this context.
Many different processes leading to $u \bar u$, $d \bar d$
or $s \bar s$ final states have been considered. 
We have done also similar calculations at
collinear NLO tree-level partonic approach. 
Some regularization procedure
has been used in both cases. Both the $k_t$-factorization and 
the collinear NLO approaches lead to rather similar results.

We have shown that the results for color-singlet fusion,
constrained by the $\gamma \gamma \to f_0(980)$ form factor at 
the on-shell point $|F_{TT}(0,0)|$, 
strongly depends on the assumed flavor structure of $f_0(980)$.
Assuming $u \bar u + d \bar d$ flavor structure leads
to negligible contribution of the color-singlet 
gluon-gluon fusion mechanism.
Assuming $s \bar s$ flavor structure (more realistic in our opinion)
gives the contribution which may be important at $p_t >$ 3 GeV.
In contrast, the color evaporation mechanism is much less sensitive
to the flavor structure.

We conclude that the color-singlet gluon-gluon fusion is not able
to describe the preliminary ALICE data \cite{Lee:thesis}
in the whole range of transverse momenta.
The color evaporation model nicely describes 
the shape of transverse momentum distribution. 
To describe absolute normalization rather maximal
probabilities $(\rm{P_{CEM}} = 1)$ must be used. 
It seems too early to draw definite conclusion. 
More global picture may arise by analysis of production
of other isoscalar mesons (such as $\eta, \eta', f_2(1270),
f_1(1285)$, etc.). This clearly goes beyond the scope of the present
analysis.

We have calculated also $\Pom \Pom \to f_0(980)$ fusion contribution 
and found nonnegligible but small contribution.
This contribution is concentrated at rather small $f_0(980)$
transverse momenta ($p_t <$ 2 GeV) but its role is rather marginal.

\section*{Acknowledgments}
A.S. is indebted to Wolfgang Sch\"afer 
for a collaboration on quarkonium
production and a discussion on related issues.
This study was partially supported by the Polish National Science Centre under grant No.
2018/31/B/ST2/03537 and by the Center for Innovation and
Transfer of Natural Sciences and Engineering Knowledge in Rzesz\'ow (Poland).


\begin{thebibliography}{99}

\bibitem{LMS_2020}
P.~Lebiedowicz, R.~Maciu{\l}a, and A.~Szczurek,
\newblock {a paper in preparation}.

\bibitem{Lee:thesis}
G.~R. Lee, {Ph.D. thesis, \textit{Resonance production in the $\pi^+ \pi^-$
  decay channel in proton-proton collisions at 7 TeV}, University of
  Birmingham, June 6, 2016}.
  \url{http://www.hep.ph.bham.ac.uk/publications/thesis/grl_thesis.pdf}.

\bibitem{Close:2002zu}
F.~E. Close and N.~A. T{\"o}rnqvist, {\em {Scalar mesons above and below 1 GeV},}
  \href{http://dx.doi.org/10.1088/0954-3899/28/10/201}{J. Phys. {\bfseries G28}
  (2002) R249},
\href{http://arxiv.org/abs/hep-ph/0204205}{{arXiv:hep-ph/0204205 [hep-ph]}}.

\bibitem{Maiani:2004uc}
L.~Maiani, F.~Piccinini, A.~D. Polosa, and V.~Riquer, {\em {New Look at Scalar
  Mesons},} \href{http://dx.doi.org/10.1103/PhysRevLett.93.212002}{Phys. Rev.
  Lett. {\bfseries 93} (2004) 212002},
\href{http://arxiv.org/abs/hep-ph/0407017}{{arXiv:hep-ph/0407017 [hep-ph]}}.

\bibitem{Tornqvist:1995kr}
N.~A. T{\"o}rnqvist, {\em {Understanding the scalar meson $q \bar{q}$ nonet},}
  \href{http://dx.doi.org/10.1007/BF01565264}{Z. Phys. {\bfseries C68} (1995)
  647},
\href{http://arxiv.org/abs/hep-ph/9504372}{{arXiv:hep-ph/9504372 [hep-ph]}}.

\bibitem{Boglione:2002vv}
M.~Boglione and M.~R. Pennington, {\em {Dynamical generation of scalar
  mesons},} \href{http://dx.doi.org/10.1103/PhysRevD.65.114010}{Phys. Rev.
  {\bfseries D65} (2002) 114010},
\href{http://arxiv.org/abs/hep-ph/0203149}{{arXiv:hep-ph/0203149 [hep-ph]}}.

\bibitem{Jaffe:1976ig}
R.~L. Jaffe, {\em {Multiquark hadrons. I. Phenomenology of $Q^{2}\bar{Q}^{2}$
  mesons},}
\href{http://dx.doi.org/10.1103/PhysRevD.15.267}{Phys. Rev. {\bfseries D15}
  (1977) 267}.

\bibitem{Jaffe:1976ih}
R.~L. Jaffe, {\em {Multiquark hadrons. II. Methods},}
\href{http://dx.doi.org/10.1103/PhysRevD.15.281}{Phys. Rev. {\bfseries D15}
  (1977) 281}.

\bibitem{Weinstein:1982gc}
J.~D. Weinstein and N.~Isgur, {\em {Do Multiquark Hadrons Exist?},}
\href{http://dx.doi.org/10.1103/PhysRevLett.48.659}{Phys. Rev. Lett. {\bfseries
  48} (1982) 659}.

\bibitem{Weinstein:1983gd}
J.~D. Weinstein and N.~Isgur, {\em {$q q \bar{q} \bar{q}$ system in a potential
  model},}
\href{http://dx.doi.org/10.1103/PhysRevD.27.588}{Phys. Rev. {\bfseries D27}
  (1983) 588}.

\bibitem{Weinstein:1990gu}
J.~D. Weinstein and N.~Isgur, {\em {$K \bar{K}$ molecules},}
\href{http://dx.doi.org/10.1103/PhysRevD.41.2236}{Phys. Rev. {\bfseries D41}
  (1990) 2236}.

\bibitem{Baru:2003qq}
V.~Baru, J.~Haidenbauer, C.~Hanhart, {\relax Yu}.~Kalashnikova, and A.~E.
  Kudryavtsev, {\em {Evidence that the $a_{0}(980)$ and $f_{0}(980)$ are not
  elementary particles},}
  \href{http://dx.doi.org/10.1016/j.physletb.2004.01.088}{Phys. Lett.
  {\bfseries B586} (2004) 53},
\href{http://arxiv.org/abs/hep-ph/0308129}{{arXiv:hep-ph/0308129 [hep-ph]}}.

\bibitem{Hooft:2008we}
G.~'t~Hooft, G.~Isidori, L.~Maiani, A.~D. Polosa, and V.~Riquer, {\em {A theory
  of scalar mesons},}
  \href{http://dx.doi.org/10.1016/j.physletb.2008.03.036}{Phys. Lett.
  {\bfseries B662} (2008) 424},
\href{http://arxiv.org/abs/0801.2288}{{arXiv:0801.2288 [hep-ph]}}.

\bibitem{Fleischer:2011au}
R.~Fleischer, R.~Knegjens, and G.~Ricciardi, {\em {Anatomy of $B^0_{s,d} \to
  J/\psi f_0(980)$},}
  \href{http://dx.doi.org/10.1140/epjc/s10052-011-1832-x}{Eur. Phys. J.
  {\bfseries C71} (2011) 1832},
\href{http://arxiv.org/abs/1109.1112}{{arXiv:1109.1112 [hep-ph]}}.

\bibitem{Maiani:2007iw}
L.~Maiani, A.~D. Polosa, and V.~Riquer, {\em {Structure of light scalar mesons
  from $D_{s}$ and $D^{0}$ non-leptonic decays},}
  \href{http://dx.doi.org/10.1016/j.physletb.2007.05.051}{Phys. Lett.
  {\bfseries B651} (2007) 129},
\href{http://arxiv.org/abs/hep-ph/0703272}{{arXiv:hep-ph/0703272 [hep-ph]}}.

\bibitem{Cheng:2005nb}
H.-Y. Cheng, C.-K. Chua, and K.-C. Yang, {\em {Charmless hadronic $B$ decays
  involving scalar mesons: Implications to the nature of light scalar mesons},}
  \href{http://dx.doi.org/10.1103/PhysRevD.73.014017}{Phys. Rev. {\bfseries
  D73} (2006) 014017},
\href{http://arxiv.org/abs/hep-ph/0508104}{{arXiv:hep-ph/0508104 [hep-ph]}}.

\bibitem{Stone:2013eaa}
S.~Stone and L.~Zhang, {\em {Use of $B\to J/\psi f_0$ Decays to Discern the $q
  \bar{q}$ or Tetraquark Nature of Scalar Mesons},}
  \href{http://dx.doi.org/10.1103/PhysRevLett.111.062001}{Phys. Rev. Lett.
  {\bfseries 111} no.~6, (2013) 062001},
\href{http://arxiv.org/abs/1305.6554}{{arXiv:1305.6554 [hep-ex]}}.

\bibitem{Tanabashi:2018oca}
M.~Tanabashi {\em et~al.}, (Particle Data Group), {\em {Review of Particle
  Physics},}
\href{http://dx.doi.org/10.1103/PhysRevD.98.030001}{Phys. Rev. {\bfseries D98}
  no.~3, (2018) 030001}.

\bibitem{Aaij:2014siy}
R.~Aaij {\em et~al.}, (LHCb Collaboration), {\em {Measurement of the resonant
  and CP components in $\overline{B}^0\to J/\psi \pi^+\pi^-$ decays},}
  \href{http://dx.doi.org/10.1103/PhysRevD.90.012003}{Phys. Rev. {\bfseries
  D90} no.~1, (2014) 012003},
\href{http://arxiv.org/abs/1404.5673}{{arXiv:1404.5673 [hep-ex]}}.

\bibitem{Pauk:2014rta}
V.~Pauk and M.~Vanderhaeghen, {\em {Single meson contributions to the muon`s
  anomalous magnetic moment},}
  \href{http://dx.doi.org/10.1140/epjc/s10052-014-3008-y}{Eur. Phys. J.
  {\bfseries C74} no.~8, (2014) 3008},
\href{http://arxiv.org/abs/1401.0832}{{arXiv:1401.0832 [hep-ph]}}.

\bibitem{Colangelo:2014dfa}
G.~Colangelo, M.~Hoferichter, M.~Procura, and P.~Stoffer, {\em {Dispersive
  approach to hadronic light-by-light scattering},}
  \href{http://dx.doi.org/10.1007/JHEP09(2014)091}{JHEP {\bfseries 09} (2014)
  091},
\href{http://arxiv.org/abs/1402.7081}{{arXiv:1402.7081 [hep-ph]}}.

\bibitem{Dorokhov:2015psa}
A.~E. Dorokhov, A.~E. Radzhabov, and A.~S. Zhevlakov, {\em {Dynamical quark
  loop light-by-light contribution to muon g-2 within the nonlocal chiral quark
  model},} \href{http://dx.doi.org/10.1140/epjc/s10052-015-3577-4}{Eur. Phys.
  J. {\bfseries C75} no.~9, (2015) 417},
\href{http://arxiv.org/abs/1502.04487}{{arXiv:1502.04487 [hep-ph]}}.

\bibitem{DeFazio:2001uc}
F.~De~Fazio and M.~R. Pennington, {\em {Probing the structure of $f_{0}(980)$
  through radiative $\phi$ decays},}
  \href{http://dx.doi.org/10.1016/S0370-2693(01)01200-X}{Phys. Lett. {\bfseries
  B521} (2001) 15},
\href{http://arxiv.org/abs/hep-ph/0104289}{{arXiv:hep-ph/0104289 [hep-ph]}}.

\bibitem{Kroll:2016mbt}
P.~Kroll, {\em {A study of the $\gamma^* - f_{0}(980)$ transition form
  factors},} \href{http://dx.doi.org/10.1140/epjc/s10052-017-4661-8}{Eur. Phys.
  J. {\bfseries C77} no.~2, (2017) 95},
\href{http://arxiv.org/abs/1610.01020}{{arXiv:1610.01020 [hep-ph]}}.

\bibitem{Pascalutsa:2012pr}
V.~Pascalutsa, V.~Pauk, and M.~Vanderhaeghen, {\em {Light-by-light scattering
  sum rules constraining meson transition form factors},}
  \href{http://dx.doi.org/10.1103/PhysRevD.85.116001}{Phys. Rev. {\bfseries
  D85} (2012) 116001},
\href{http://arxiv.org/abs/1204.0740}{{arXiv:1204.0740 [hep-ph]}}.

\bibitem{Babiarz:2020jkh}
I.~Babiarz, R.~Pasechnik, W.~Sch{\"a}fer, and A.~Szczurek, {\em
  {Hadroproduction of scalar $P$-wave quarkonia in the light-front
  $k_T$-factorization approach},}
\href{http://arxiv.org/abs/2002.09352}{{arXiv:2002.09352 [hep-ph]}}.

\bibitem{Cisek:2017gno}
A.~Cisek and A.~Szczurek, {\em {Prompt inclusive production of $J/\psi$,
  $\psi'$ and $\chi_{c}$ mesons at the LHC in forward directions within the
  NRQCD $k_t$-factorization approach: Search for the onset of gluon
  saturation},} \href{http://dx.doi.org/10.1103/PhysRevD.97.034035}{Phys. Rev.
  {\bfseries D97} no.~3, (2018) 034035},
\href{http://arxiv.org/abs/1712.07943}{{arXiv:1712.07943 [hep-ph]}}.

\bibitem{Babiarz:2019mag}
I.~Babiarz, R.~Pasechnik, W.~Sch{\"a}fer, and A.~Szczurek, {\em {Prompt
  hadroproduction of $\eta_c(1S,2S)$ in the $k_T$-factorization approach},}
  \href{http://dx.doi.org/10.1007/JHEP02(2020)037}{JHEP {\bfseries 02} (2020)
  037},
\href{http://arxiv.org/abs/1911.03403}{{arXiv:1911.03403 [hep-ph]}}.

\bibitem{Fritzsch:1977ay}
H.~Fritzsch, {\em {Producing heavy quark flavors in hadronic collisions: A test
  of quantum chromodynamics},}
\href{http://dx.doi.org/10.1016/0370-2693(77)90108-3}{Phys. Lett. {\bfseries
  B67} (1977) 217}.

\bibitem{Halzen:1977rs}
F.~Halzen, {\em {CVC for gluons and hadroproduction of quark flavors},}
\href{http://dx.doi.org/10.1016/0370-2693(77)90144-7}{Phys. Lett. {\bfseries
  B69} (1977) 105}.

\bibitem{Maciula:2018bex}
R.~Maciu{\l}a, A.~Szczurek, and A.~Cisek, {\em {$J/\psi$-meson production
  within improved color evaporation model with the $k_{T}$-factorization
  approach for $c\bar{c}$ production},}
  \href{http://dx.doi.org/10.1103/PhysRevD.99.054014}{Phys. Rev. {\bfseries
  D99} no.~5, (2019) 054014},
\href{http://arxiv.org/abs/1810.08063}{{arXiv:1810.08063 [hep-ph]}}.

\bibitem{Antchev:2013gaa}
G.~Antchev {\em et~al.}, (TOTEM Collaboration), {\em {Measurement of
  proton-proton elastic scattering and total cross-section at $\sqrt{s}=7$ TeV},}
\href{http://dx.doi.org/10.1209/0295-5075/101/21002}{EPL {\bfseries 101} no.~2,
  (2013) 21002}.

\bibitem{Aad:2014dca}
G.~Aad {\em et~al.}, (ATLAS Collaboration), {\em {Measurement of the total
  cross section from elastic scattering in pp collisions at $\sqrt{s}=7$ TeV
  with the ATLAS detector},}
  \href{http://dx.doi.org/10.1016/j.nuclphysb.2014.10.019}{Nucl. Phys.
  {\bfseries B889} (2014) 486},
\href{http://arxiv.org/abs/1408.5778}{{arXiv:1408.5778 [hep-ex]}}.

\bibitem{Jung:2010si}
H.~Jung {\em et~al.}, {\em {The CCFM Monte Carlo generator CASCADE version
  2.2.03},} \href{http://dx.doi.org/10.1140/epjc/s10052-010-1507-z}{Eur. Phys.
  J. {\bfseries C70} (2010) 1237},
\href{http://arxiv.org/abs/1008.0152}{{arXiv:1008.0152 [hep-ph]}}.

\bibitem{Kimber:2001sc}
M.~A. Kimber, A.~D. Martin, and M.~G. Ryskin, {\em {Unintegrated parton
  distributions},} \href{http://dx.doi.org/10.1103/PhysRevD.63.114027}{Phys.
  Rev. {\bfseries D63} (2001) 114027},
\href{http://arxiv.org/abs/hep-ph/0101348}{{arXiv:hep-ph/0101348 [hep-ph]}}.

\bibitem{Watt:2003vf}
G.~Watt, A.~D. Martin, and M.~G. Ryskin, {\em {Unintegrated parton
  distributions and electroweak boson production at hadron colliders},}
  \href{http://dx.doi.org/10.1103/PhysRevD.70.014012,
  10.1103/PhysRevD.70.079902}{Phys. Rev. {\bfseries D70} (2004) 014012},
  \href{http://arxiv.org/abs/hep-ph/0309096}{{arXiv:hep-ph/0309096 [hep-ph]}}.
[Erratum: Phys. Rev.D70,079902(2004)].

\bibitem{Martin:2009ii}
A.~D. Martin, M.~G. Ryskin, and G.~Watt, {\em {NLO prescription for
  unintegrated parton distributions},}
  \href{http://dx.doi.org/10.1140/epjc/s10052-010-1242-5}{Eur. Phys. J.
  {\bfseries C66} (2010) 163},
\href{http://arxiv.org/abs/0909.5529}{{arXiv:0909.5529 [hep-ph]}}.

\bibitem{Hautmann:2013tba}
F.~Hautmann and H.~Jung, {\em {Transverse momentum dependent gluon density from
  DIS precision data},}
  \href{http://dx.doi.org/10.1016/j.nuclphysb.2014.03.014}{Nucl. Phys.
  {\bfseries B883} (2014) 1},
\href{http://arxiv.org/abs/1312.7875}{{arXiv:1312.7875 [hep-ph]}}.

\bibitem{Ciafaloni:1987ur}
M.~Ciafaloni, {\em {Coherence effects in initial jets at small $Q^{2}/s$},}
\href{http://dx.doi.org/10.1016/0550-3213(88)90380-X}{Nucl. Phys. {\bfseries
  B296} (1988) 49}.

\bibitem{Catani:1989yc}
S.~Catani, F.~Fiorani, and G.~Marchesini, {\em {QCD coherence in initial state
  radiation},}
\href{http://dx.doi.org/10.1016/0370-2693(90)91938-8}{Phys. Lett. {\bfseries
  B234} (1990) 339}.

\bibitem{Catani:1989sg}
S.~Catani, F.~Fiorani, and G.~Marchesini, {\em {Small-$x$ behavior of initial
  state radiation in perturbative QCD},}
\href{http://dx.doi.org/10.1016/0550-3213(90)90342-B}{Nucl. Phys. {\bfseries
  B336} (1990) 18}.

\bibitem{Ewerz:2013kda}
C.~Ewerz, M.~Maniatis, and O.~Nachtmann, {\em {A Model for Soft High-Energy
  Scattering: Tensor Pomeron and Vector Odderon},}
  \href{http://dx.doi.org/http://dx.doi.org/10.1016/j.aop.2013.12.001}{Annals
  Phys. {\bfseries 342} (2014) 31},
\href{http://arxiv.org/abs/1309.3478}{{arXiv:1309.3478 [hep-ph]}}.

\bibitem{Lebiedowicz:2013ika}
P.~Lebiedowicz, O.~Nachtmann, and A.~Szczurek, {\em {Exclusive central
  diffractive production of scalar and pseudoscalar mesons; tensorial vs.
  vectorial pomeron},}
  \href{http://dx.doi.org/10.1016/j.aop.2014.02.021}{Annals Phys. {\bfseries
  344} (2014) 301},
\href{http://arxiv.org/abs/1309.3913}{{arXiv:1309.3913 [hep-ph]}}.

\bibitem{Lebiedowicz:2016ioh}
P.~Lebiedowicz, O.~Nachtmann, and A.~Szczurek, {\em {Central exclusive
  diffractive production of the $\pi^{+}\pi^{-}$ continuum, scalar, and tensor
  resonances in $pp$ and $p \bar{p}$ scattering within the tensor Pomeron
  approach},} \href{http://dx.doi.org/10.1103/PhysRevD.93.054015}{Phys. Rev.
  {\bfseries D93} (2016) 054015},
\href{http://arxiv.org/abs/1601.04537}{{arXiv:1601.04537 [hep-ph]}}.

\bibitem{Lebiedowicz:2018eui}
P.~Lebiedowicz, O.~Nachtmann, and A.~Szczurek, {\em {Towards a complete study
  of central exclusive production of $K^{+}K^{-}$ pairs in proton-proton
  collisions within the tensor Pomeron approach},}
  \href{http://dx.doi.org/10.1103/PhysRevD.98.014001}{Phys. Rev. {\bfseries
  D98} (2018) 014001},
\href{http://arxiv.org/abs/1804.04706}{{arXiv:1804.04706 [hep-ph]}}.

\bibitem{CMS:pippim}
A.~M. Sirunyan {\em et~al.}, (CMS Collaboration), {\em {Study of central
  exclusive $\pi^+\pi^-$ production in proton-proton collisions at $\sqrt{s} =
  5.02$ and 13 TeV},} CMS-FSQ-16-006, CERN-EP-2020-005,
\href{http://arxiv.org/abs/2003.02811}{{arXiv:2003.02811 [hep-ex]}}.

\bibitem{GolecBiernat:2001mm}
K.~J. Golec-Biernat and M.~Wusthoff, {\em {Diffractive parton distributions
  from the saturation model},}
  \href{http://dx.doi.org/10.1007/s100520100661}{Eur. Phys. J. {\bfseries C20}
  (2001) 313},
\href{http://arxiv.org/abs/hep-ph/0102093}{{arXiv:hep-ph/0102093 [hep-ph]}}.

\bibitem{Kutak:2014wga}
K.~Kutak, {\em {Hard scale dependent gluon density, saturation and
  forward-forward dijet production at the LHC},}
  \href{http://dx.doi.org/10.1103/PhysRevD.91.034021}{Phys. Rev. {\bfseries
  D91} (2015) 034021},
\href{http://arxiv.org/abs/1409.3822}{{arXiv:1409.3822 [hep-ph]}}.

\bibitem{Gluck:2008gs}
M.~Gl{\"u}ck, P.~Jimenez-Delgado, E.~Reya, and C.~Schuck, {\em {On the role of
  heavy flavor parton distributions at high energy colliders},}
  \href{http://dx.doi.org/10.1016/j.physletb.2008.04.063}{Phys.Lett. {\bfseries
  B664} (2008) 133},
\href{http://arxiv.org/abs/0801.3618}{{arXiv:0801.3618 [hep-ph]}}.

\bibitem{vanHameren:2016kkz}
A.~van Hameren, {\em {KaTie: For parton-level event generation with
  $k_T$-dependent initial states},}
  \href{http://dx.doi.org/10.1016/j.cpc.2017.11.005}{Comput. Phys. Commun.
  {\bfseries 224} (2018) 371},
\href{http://arxiv.org/abs/1611.00680}{{arXiv:1611.00680 [hep-ph]}}.

\bibitem{Dulat:2015mca}
S.~Dulat, T.-J. Hou, J.~Gao, M.~Guzzi, J.~Huston, P.~Nadolsky, J.~Pumplin,
  C.~Schmidt, D.~Stump, and C.~P. Yuan, {\em {New parton distribution functions
  from a global analysis of quantum chromodynamics},}
  \href{http://dx.doi.org/10.1103/PhysRevD.93.033006}{Phys. Rev. {\bfseries
  D93} no.~3, (2016) 033006},
\href{http://arxiv.org/abs/1506.07443}{{arXiv:1506.07443 [hep-ph]}}.

\bibitem{Harland-Lang:2014zoa}
L.~A. Harland-Lang, A.~D. Martin, P.~Motylinski, and R.~S. Thorne, {\em {Parton
  distributions in the LHC era: MMHT 2014 PDFs},}
  \href{http://dx.doi.org/10.1140/epjc/s10052-015-3397-6}{Eur. Phys. J.
  {\bfseries C75} no.~5, (2015) 204},
\href{http://arxiv.org/abs/1412.3989}{{arXiv:1412.3989 [hep-ph]}}.

\bibitem{Kutak:2016mik}
K.~Kutak, R.~Maciula, M.~Serino, A.~Szczurek and A.~van Hameren,
{\em {Four-jet production in single- and double-parton scattering within high-energy factorization},}
  \href{http://dx.doi.org/10.1007/JHEP04(2016)175}{JHEP {\bfseries 04} (2016)
  175},
\href{http://arxiv.org/abs/1602.06814}{{arXiv:1602.06814 [hep-ph]}}.

\bibitem{Mori:2006jj}
T.~Mori {\em et~al.}, (Belle Collaboration), {\em {High statistics study of the
  $f_{0}(980)$ resonance in $\gamma \gamma \to \pi^{+} \pi^{-}$ production},}
  \href{http://dx.doi.org/10.1103/PhysRevD.75.051101}{Phys. Rev. {\bfseries
  D75} (2007) 051101},
\href{http://arxiv.org/abs/hep-ex/0610038}{{arXiv:hep-ex/0610038 [hep-ex]}}.

\bibitem{Uehara:2008ep}
S.~Uehara {\em et~al.}, (Belle Collaboration), {\em {High-statistics
  measurement of neutral-pion pair production in two-photon collisions},}
  \href{http://dx.doi.org/10.1103/PhysRevD.78.052004}{Phys. Rev. {\bfseries
  D78} (2008) 052004},
\href{http://arxiv.org/abs/0805.3387}{{arXiv:0805.3387 [hep-ex]}}.

\end{thebibliography}

\end{document}